\documentclass[a4paper, 12pt]{article}
\usepackage[T1]{fontenc}
\usepackage[utf8]{inputenc}
\usepackage{amsfonts}
\usepackage{amssymb}
\usepackage{indentfirst}
\usepackage{amsmath}
\usepackage{amsthm}
\usepackage[english]{babel}
\usepackage[pdftex]{graphicx}
\newtheorem{definicja}{Definition}
\newtheorem{twierdzenie}{Theorem}
\newtheorem{uwaga}{Remark}
\newtheorem{fakt}{Fact}

\begin{document}
\noindent
\begin{center}
\Large
UNIVERSITY OF SZCZECIN\\
MATHEMATICS AND PHYSICS DEPARTMENT\\
INSTITUTE OF MATHEMATICS
\end{center}

\vfill
\begin{center}
\Huge \bfseries
Ehreshmann theory of connection in a principal bundle - compendium for physicists.
\end{center}

\vfill
\begin{center}
\huge \mdseries
Marta Dudek\\
\large
e-mail: marta.dudek@vp.pl\\
\huge \mdseries
Janusz Garecki\\
\large
e-mail: janusz.garecki@usz.edu.pl
\end{center}

\vfill
\begin{center}
\large
Institute of Mathematics\\
University of Szczecin\\
Wielkopolska 17, 70-451 Szczecin\\
Poland, EU
\end{center}

\vfill
\begin{center}
\large
SZCZECIN 2018
\end{center}
\cleardoublepage

\newpage
\begin{center}\section{Global theory of connection in  principal fibre bundles (compendium).} \end{center}

Differential geometry and tensor analysis give main mathematical tools for relativists. But most of them use, up to now, the old index formalism in local coordinate maps on the spacetime manifold [1,2,3].

This formalism was developed in the past (about 100 years ago) by Italian mathematicians Georgio Ricci Curbastro, Tullio Levi-Civita and Luigi Bianchi and it is exhaustively presented e.g. in [4,5,6,7].

The modern coordinate-free formulation of the formalism created by Charles Ehreshmann is still not sufficiently known for majority of relativists.

We would like to give a compendium about these two formulations and connection between them. We will start with foundations of the global Ehresmann theory on fibre bundles and end up with old local index formulation on basic manifold.

Our lecture is founded on standard books [8,9,10,11].

We have assumed that a potential reader knows elements of differentiable manifolds and Lie groups.

\newpage
\begin{center}\subsection{Fibre bundles.} \end{center}

\begin{definicja}
Let $\mathcal{G}$ be a Lie group. A $\mathcal{G}$-space is a differentiable manifold $(M, A_{M})$ and the group $\mathcal{G}$  acts as a group of point mappings, i.e. $\mathcal{G}$-space is a pair $[M, \mathcal{G}]$, and there is a smooth mapping\\
$$\psi : M\times \mathcal{G}\rightarrow M$$
such that $\psi_{a}\circ \psi_{b}=\psi_{ba}$ and $\psi_{e}=idM$,\\
where $\psi_{a}(p)=\psi(p, a):=pa$, $p\in M$, $a, b \in \mathcal{G}$, and $e$ is a unit element of the group $\mathcal{G}$.
\end{definicja}

Here $A_{M}$ means the maximal atlas on $M$.

\begin{uwaga} By $\psi_{a}(p)=pa$ we mean, that $\mathcal{G}$ acts on the right and we denote it by $R_{a}$.
\end{uwaga}

\begin{definicja}
An orbit of the point $p\in M$ is the set of points of the manifold defined as follows
$$\lbrace R_{a}(p): a \in \mathcal{G} \rbrace$$
and it is denoted by $p\mathcal{G}$.
\end{definicja}

If $p\mathcal{G}=M$, then $M$ is called a homogenous space and we say, that $\mathcal{G}$ acts on $M$ transitively. It follows, that for any pair of points $p_{1}, p_{2} \in M$ there is $g \in \mathcal{G}$ such that $R_{g}p_{1}=p_{2}$.

\begin{definicja}
A group of isotropy (stabilizer) of the point $p\in M$ denoted by $\mathcal{G}_{p}$ is a set 
$$\mathcal{G}_{p}:=\lbrace a\in \mathcal{G}: R_{a}(p)=p \rbrace.$$ We say, that $p$ is a fixed point of the mapping $R_{a}$ and $\mathcal{G}_{p}\subset \mathcal{G}$.
\end{definicja}

Notice, that a group $\mathcal{G}_{p}$ is a closed subgroup of the group $\mathcal{G}$.

\begin{definicja}
If $\mathcal{G}_{p}=\lbrace e \rbrace$ for any point $p\in M$, then we say, that $\mathcal{G}$ acts freely on $M$ (i.e. without fixed points). Then $M$ is called a main space.
\end{definicja}

If $e\in \mathcal{G}$ is the only element of the group $\mathcal{G}$ for which $R_{g}$ is an identity, i.e. $R_{g}x=x$ for every $x\in M$, then we say that the Lie group $\mathcal{G}$ acts on $M$ effectively.

\begin{definicja}
A Lie group $\mathcal{G}$ acts simply transitive or single transitive on $M$, if for any pair of points $p_{1}, p_{2}\in M$ there is   only one point $g\in \mathcal{G}$ such that $R_{g}(p_{1})=p_{2}$.
\end{definicja}

If the Lie group $\mathcal{G}$ acts on $M$ on the right, it assignes for every vector field $A \in \mathfrak{g}$ a vector field $A^{\ast }$ on $M$ in the following way:\\
$$A \leftrightarrow a_{t}=exp(tA)\ ,$$
where $a_{t}$ is a 1-parameter group of global transformations of a manifold $M$. It is a subgroup of the group $\mathcal{G}$. The action of the group $a_{t}$ on $M$ induces on $M$ a vector field tangent to orbits $a_{t}(p),\ p \in M,\ t \in \mathbb{R}$. It is a vector field $A^{\ast } \in \mathcal{X}(M)$. $\mathcal{X}(M)$ denotes here a Lie algebra of vector fields on the manifold. $\mathfrak{g}$ stands for the group algebra $\mathcal{G}$.\\

A mapping $\sigma : \mathfrak{g} \rightarrow \mathcal{X}(M)$, which for a field $A \in \mathfrak{g}$ assigns a vector field $A^{\ast } \in \mathcal{X}(M)$, is a homomorphism of Lie algebras. If the group $\mathcal{G}$ acts on $M$ effectively, then  $\sigma : \mathfrak{g} \rightarrow \mathcal{X}(M)$ is a monomorphism. 

\begin{definicja}
A mapping $\phi $ from the algebra $\mathfrak{g}$ onto the algebra $\mathfrak{h }$ is an isomorphism, if:\\
\begin{enumerate}
\item $\phi $ is an isomorphism of vector spaces $\mathfrak{g}$ i $\mathfrak{h }$;
\item $\phi ([u, v])=[\phi (u),\phi (v)]$ for any $u,\ v \in \mathfrak{g}$.
\end{enumerate}
\end{definicja}

In the above definition $[u,v]$ is a Lie bracket in $\mathfrak{g}$, and $[\phi (u),\phi (v)]$ is a bracket of images of $u$ and $v $ in $\mathfrak{h}$. An isomorphism of the algebra $\mathfrak{g}$ onto itself is called an automorphism.

One can introduce some notions connected with fibre bundle by considering graf $\tilde {f}$ of the mapping $f:M\rightarrow N$, where $M,\ N$ are two differential manifolds.

\begin{figure}[bh]
\centering
\includegraphics[width=0.8\textwidth]{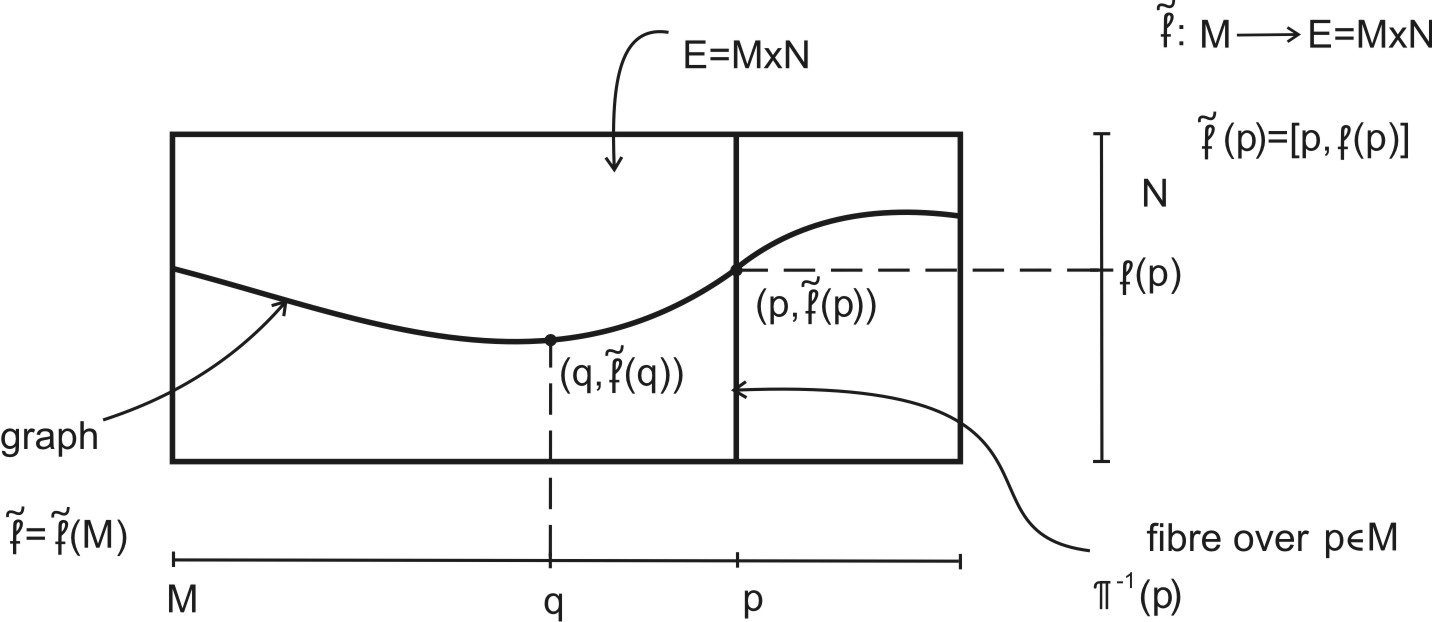}
\caption{A trivial fibre bundle.}
\end{figure}

The above picture shows a trivial fibre bundle connected with graph $\tilde {f}$ of the mapping $f: M\rightarrow N$.\\
$E=M\times N$ is called the space of the fibre or just simply the fibre over  $M$;\\
$M$ is a base space;\\
$\pi :E\rightarrow M$ is defined as $\pi (p, q):=p$ and it is called a projection (a projection of $E$ onto the base $M$), $p\in M$, $q\in N$, $(p,q)\in E$;\\
$N$ is a typical fibre;\\
$E_{p}:=\pi ^{-1}(p) \in E$ is a fibre over $p \in M$;\\
$\widetilde{f}: M\rightarrow E$ is a section of the bundle $E$ called graf of the mapping $f:\ M\rightarrow N$;
$\widetilde{f}(p)=[p, f(p)]\in M\times N=E$.

In general $E(M,\pi ,N)$ is a fibre bundle over $M$ with a typical fibre $N$ if the projection $\pi :E \rightarrow M$ is a smooth surjection of the differentiable manifold E onto the manifold M and if each point $p \in M$ has a neighborhood $U$, such, that there is a diffeomorphism $h: \pi^{-1}(U) \rightarrow U \times N$ such that $\pi \circ [h^{-1}(p, y)]=p$, $p \in U$, $y \in N$. This diffeomorphism $h: \pi^{-1}(U) \rightarrow U \times N$ is called a local trivialisation of the bundle, and it means, that a piece of $E$ over a sufficiently small $U \subset M$ looks like a product manifold $U \times N$.

\begin{definicja}
The bundle $E$ over $M$ is called a trivial bundle if there is a diffeomorphism $h : E \rightarrow M \times N$ such that $\pi[h^{-1}(p, q)]=p$ for $\forall p \in M$, $\forall q \in N$.
\end{definicja}

The bundle connected with graf $\tilde {f}$ is a trivial bundle.

Another example of the fibre bundle is a tangent bundle $T(M)$ over a differentiable manifold $M$. 

Assume that the differentiable manifold $M$ permits a global coordinate system, i.e. an atlas consisting of one chart $\varphi : M \rightarrow R^{n}$ and let us consider $T(M):=\bigcup_{p \in M} T_{p}(M)$.

In a coordinate system $(\varphi , M)$ (global coordinate system) a vector field looks as follows:
$$X=X^{i}(p)\Bigl (\frac{\partial }{\partial x^{i}}\Bigr )_{p},\ p \in M.$$

Such field gives a point mapping
$$(x^{i}):\ M\rightarrow \mathbb{R}^{n}\ (=N).$$

Let $\varphi :\ M\rightarrow \mathbb{R}^{n}$ be the chart mentioned before and denote the chart $\widehat {\varphi}:\ E=T(M)\rightarrow \mathbb{R}^{2n}$ as follows:\\
Let $X\in T(M)$, then $X\in T_{p}(M)$, for some $p\in M$ and $X=X^{i}(\frac{\partial }{\partial x^{i}})_{p}.$ Set $\widehat {\varphi}(X):=[x^{1}(p),...,x^{n}(p), X^{1}(p),...,X^{n}(p)]\in \mathbb {R}^{2n}$.

$\widehat {\varphi }$ is bijective. Moreover, if both $\varphi $ and $\psi $ are $C^{k}$-compatible (global) charts on $M$, then $\widehat {\varphi }$ and $\widehat {\psi }$ are $C^{k}$-compatible on $T(M)$.

We say that the operation "$\ \ \widehat {}\ \ $" (hat) lifts an atlas $M$ onto the atlas on $T(M)$. Lifted from $M$ onto $T(M)$ the atlas with "hat" changes $T(M)$ into $2n$-dimensional differentiable manifold, which is called a tangent bundle of the differentiable manifold $M$. (In our example with one global chart this bundle is trivial).

\begin{figure}[h]
\centering
\includegraphics [width=0.8\textwidth]{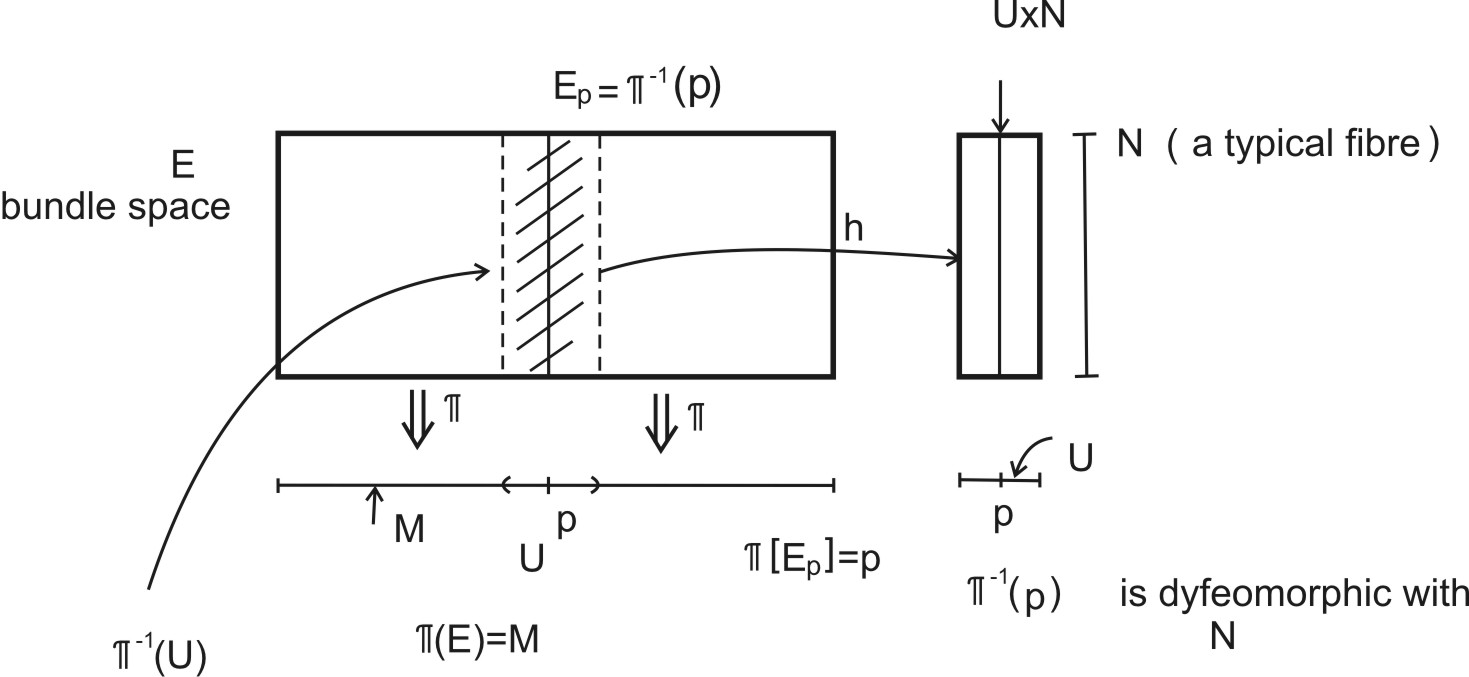}
\end{figure}

\begin{figure}[h]
\centering
\includegraphics [width=0.8\textwidth]{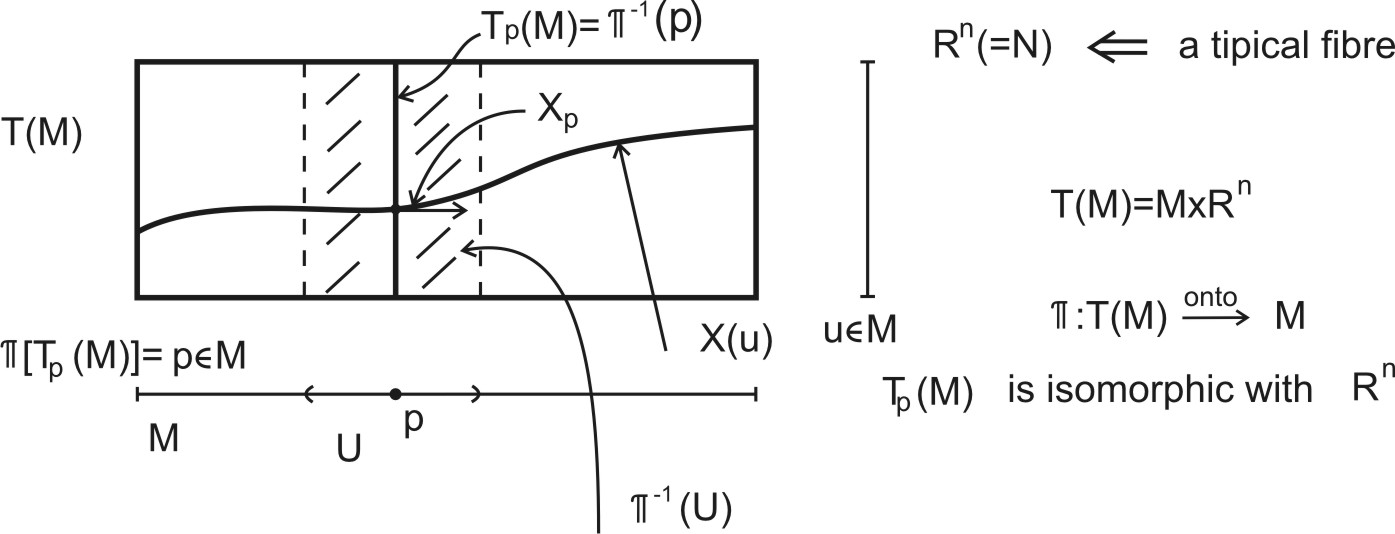}
\caption{Pictures of the fibre bundles $E(M,\pi ,N)$ and $T(M)$.}
\end{figure}

\begin{center}\subsection{Principal fibre bundles.} \end{center}

In the present section we present a definition of principal fibre bundle and some examples of this structure.  These bundles play very important role in the differential geometry and they are applied in physics for example in the theory of gauge fields.

\begin{definicja}
Let $M$ be a manifold and $G$ a Lie group. A (differentiable) principal fibre bundle over $M$ with group $G$ consists of a manifold $P$ and an action of $G$ on $P$ satisfying the following conditions:
\begin{enumerate}
\item $G$ acts freely on $P$ on the right; $(u, a)\in P\times G\rightarrow ua:=R_{a}u \in P$; 
\item $M$ is the quotient space of $P$ by the equivalence relation induced by $G$, $M=P/G$, and the canonical projection $\pi:\ P\rightarrow M$ is differentiable;

The action of $G$ on $P$ introduces on $P$ the equivalence relation: $u_{1}\approx u_{2}\iff u_{2}=R_{a}u_{1}$, which splits $P$ into the sum of the separable equivalence classes (fibres). The set of these equivalence classes is denoted by $P/G$ and is diffeomorphic with $M$.

\item $P$ is locally trivial, that is, every point $x \in M$ has a neighborhood $U$ such that $\pi ^{-1}(U)$ is isomorphic with $U \times G$ in the sense that there is a diffeomorphism $\psi :\pi^{-1}(U)\rightarrow U\times G$ such that $\psi (u)=[\pi (u), \phi (u)]$, where $\phi (u):\pi^{-1}(U)\rightarrow G$ is defined as follows $\phi (ua)=(\phi(u))a$ for every $u \in \pi^{-1}(U)$ and $a \in G.$
\end{enumerate}
\end{definicja}

A principal fibre bundle is denoted by $P(M, G, \pi )$ or $P(M, G)$.

\begin{definicja}
The differential manifold $P$ is called the bundle space; the manifold $M$ is the base space or the base of principal bundle $P$; $G$ is the structure group of bundle $P$ and the mapping $\pi :P \rightarrow M$ projection.
\end{definicja}

\begin{uwaga}
\begin{enumerate}
\item For each $x \in M$ $\pi ^{-1}(x)$ is a closed submanifold of $P$, called the fibre over $x$. If $u \in \pi^{-1}(x)$, then the fibre $\pi ^{-1}(x)$ is the set of points $ua$, $a\in G$ and is called the fibre through $u$.
\item Every fibre is diffeomorphic to the structure group $G$.
\item Given a Lie group $G$ and a manifold $M$, $G$ acts freely on $P=M\times G$ on the right as follows. For each $b\in G$ $R_{b}$ maps $(x, a)\in M\times G$ into $(x, ab) \in M\times G$. The principal fibre bundle $P(M, G, \pi )$ thus obtained is trivial. 
\item From local triviality of $P(M, G)$ we see, that if $W$ is a submanifold of $M$ then $\pi ^{-1}(W)(W, G)\subset P$ is a principal fibre bundle. The bundle $\pi ^{-1}(W)(W, G)$ is called the portion of $P$ over $W$ or the restriction of $P$ to $W$ or a piece of the bundle $P$ over $W$. 
\end{enumerate}
\end{uwaga}

Let $P(M,G)$ be a principal fibre bundle. The action of the group $G$ on $P$ induces a homomorphism $\sigma ^{*}:\mathfrak{g}\rightarrow \chi (P)$. Here $\mathfrak{g}$ is the Lie algebra of $G$ and $\chi (P)$ is the Lie algebra of vector fields on $P$.

For each $u\in P$ let $\sigma _{u}$ be the mapping $a\in G\rightarrow u\cdot a\in P$ ($ \sigma _{u}:\ G\rightarrow P$, $\sigma _{u} (a)=ua \in P$, $u \in P$), then $(\sigma _{u})^{*}$ $A_{e}=(\sigma ^{*}A)_{u}$. Tangent mapping $\sigma ^{*}$ is a linear mapping of $\mathfrak{g}$ into $\chi (P)$:
$$(\sigma _{u})^{*}:T_{e}(G)\rightarrow T_{u}(P)\subset \chi (P).$$
The algebra $\mathfrak{g}$ is here identified with the tangent space $T_{e}(G)$ in unit element $e\in G$. $A_{e}=A(e)\in T_{e}(G)$, where $A\in \mathfrak{g}$.

\begin{definicja}
For each $A\in \mathfrak{g}$ the field $A^{*}=\sigma ^{*}(A)$ is called the fundamental vector field corresponding to $A\in \mathfrak{g}.$
\end{definicja}
Since $G$ acts (vertically) on fibres: $G$ maps each fibre into itself, so the vector $A^{*}_{u}$ is tangent to the fibre at each $u\in P$. 

Because the dimension of each fibre is equal to that of $dim\ \mathfrak{g}$, the mapping $\mathfrak{g}\ni A\rightarrow (A^{*})_{u}$ of the algebra $\mathfrak{g}$ into $T^{v}_{u}(P)$ is a linear isomorphism of $\mathfrak{g}$ onto the vector space $T^{v}_{u}(P)$ tangent to the fibre at $u\in P$ and called the tangent space of vertical vectors.

\begin{fakt}
Let $A^{*}$ be the fundamental vector field corresponding to $A\in \mathfrak{g}$. Then, for each $a \in G$, the vector field $R_{a}^{*}(A^{*})$ is the fundamental vector field over $P$, corresponding to the field $Ad_{a^{-1}}A\in \mathfrak{g}$. $[(Ad_{a^{-1}})A=(R_{a})^{*}\cdot L^{*}_{a^{-1}}(A)=(R_{a})^{*}A]$
\end{fakt}

Here $Ad$ denotes an adjoint representation of the group $G$ in its own algebra $\mathfrak{g}$.

\begin{uwaga}
Fundamental vector fields are important in the theory of connection. 
\end{uwaga}

Here we present some examples of principal fibre bundles:
\begin{enumerate}
\item The bundle of linear frames $L(M)$ with the structure group $GL(n, \mathbb{R})$.\\
Let $(M, A_{M})$ be a manifold set on $\mathbb{R}^{n}$, where $A_{M}$ is a maximal atlas. A linear frame $u$ at a point $x \in M$ is an ordered basis $(X_{1},..., X_{n})$ of the tangent space $T_{x}(M)$.\\
 Let $L(M)=\bigcup_{x \in M}$ $\lbrace $the set of all linear basis $u$ at $x \in M \rbrace $ and let define $\pi : L(M) \rightarrow M$ in the following way: $\pi[u(x)]=x \in M$, where $u(x)$ is a basis $T_{x}(M)$. The general linear group $GL(n, \mathbb{R})$ acts on $L(M)$ on the right as follows: if $a=(a^{i}_{\ j} \in GL(n, \mathbb{R})$ and $u=(X_{1},...,X_{n})$ is a linear frame at $x \in M$, then $ua:=(Y_{1},...,Y_{n})$, where $Y_{i}=X_{k}a^{k}_{\ i}$ is the new linear frame $u^{'}$ at the point $x \in M$. The group $GL(n, \mathbb{R})$ acts freely on $L(M)$. Moreover $\pi(u)=\pi(v)$ if and only if  $v=ua$ for some $a \in GL(n, \mathbb{R})$. 

\underline{Differentialiable structure on $L(M)$}\\
Let $(x^{1},...,x^{n})$ be a local coordinate system in a coordinate neighborhood $U \subset M$. Every frame $u$ at $x \in U$ can be expressed uniquely in the form $u=(X_{1},...,X_{n})$ with $X_{i}=X_{i}^{\ k}(\frac{\partial }{\partial x^{k}})_{x}$ and $det[X^{\ k}_{i}]\neq 0$. $\pi ^{-1}(U)$ is diffeomorphic with $U\times GL(n,\mathbb{R})$. We can make $L(M)$ into a differentiable manifold by taking $(x^{i})$ and $(X^{\ k}_{i})$ as a local coordinate system in $\pi ^{-1}(U)$. [$(x^{i})=$ local coordinates on $U\subset M$, a $X^{\ k}_{i}=$ coordinates (components) of the frame $u\in T_{x}(M)$ in a natural frame $\lbrace (\frac{\partial }{\partial x^{i}})_{x}$ given by a local map $(\varphi , U)$. This map gives us local coordinates $(x^{i})$ on $U\subset M$.] $L(M)$ $[M, GL(n, \mathbb{R})]$ is the principal fibre bundle over $M$ with the structure group $GL(n,\mathbb{R})$. This bundle is called the bundle of linear frames over $M$.

\item The bundle of orthonormal frames $\mathcal{O}[M, O(n)]$.\\
Let $(M_{n}, A_{M})$ be a Riemann manifold. 
\begin{definicja}
The basis $\left \{\vec{e}_{i}\right \}(x)_{(i=1,2,...,n)}$ on $M_{n}$, $x\in M_{n}$ is called an orthonormal basis if $g_{x}(\vec{e}_{i},\vec{e}_{j})=\left \langle\vec{e}_{i}|\vec{e}_{j}\right \rangle =\delta _{ij}$ ($g$ - a metric on $M_{n}$, $g_{x}=:g(x)$).
\end{definicja}
When we have an orthonormal frame $\left \{\vec{e}_{i}\right \}(x)_{(i=1,...,n)}$ on $M_{n}$ we can obtain from it any other orthonormal frame $\left \{ \vec{e}_{i^{'}}\right \}_{(i^{'}=1,...,n)}$ with the help of transformation $\vec{e}_{i^{'}}=\vec{e}_{j}A^{j}_{i^{'}}$, where $[ A^{j}_{i^{'}}]\in O(n)\subset GL(n,\mathbb{R})$. $O(n)$ denotes here an orthonormal group. There is a natural bijection between the collection of orthonormal frames attached at a point $x\in M$ and the group $O(n)$. The sum $\bigcup _{x\in M}(x,o_{x})$, where $x\in M_{n}$ and $o_{x}$ is a collection of orthonormal frames at $x$ is denoted by $\mathcal{O}(M)$ and it is called the principal bundle of orthonormal frames. 
\begin{definicja}
The principal bundle of orthonormal frames over a Riemannian manifold $(M_{n},g)$ is the set $\mathcal{O}(M)$. This bundle we usually denote by $\mathcal{O}\bigl [M_{n},O(n), \pi \bigr ]$ or simply $\mathcal{O}(M)$.
\end{definicja}
The structure of the manifold is given by local coordinates at $\pi ^{-1}(U):(x^{i}),(A^{j}_{i^{'}})$ $(i,\ j,\ i^{'}=1,...,n)$, where $(x^{i})$- are local coordinates on $U\subset M_{n}$, a $(A^{j}_{i^{'}})\in O(n)$. 
$$\pi^{-1}(U)=U\times O(n)$$
$\mathcal{O}(M)$ is a restriction of the principal bundle $L(M)$ to the orthogonal group $O(n)$. One can write, that $\mathcal{O}(M)=\left \{ u\in L(M):g_{ij}(u)=\delta_{ij}\right \}$, $g_{ij}(u):=g(\vec{e}_{i},\vec{e}_{j})$.
\end{enumerate}

\newpage
\begin{center}\section{A connection in the fibre bundle.} \end{center}

Let $P(M, G)$ be a principal fibre bundle over a manifold $M$ with a structure group $G$. Let $T_{u}P$ be a tangent space at the point $u \in P$ and let $V_{u}$ be the subspace of the space $T_{u}P$  consisting of vectors tangent to the fibre through $u$. 

\begin{definicja}
A connection $\Gamma $ in the principal fibre bundle $P$ is an assignment of a subspace $H_{u}$ a space $T_{u}P$ at each point $u \in P$ such that:
\begin{enumerate}
\item $T_{u}P$ is a direct sum of the subspaces $V_{u}$ i $H_{u}$
$$T_{u}P=V_{u}\oplus H_{u}\ ;$$
\item $H_{ua}=(R_{a})^{* }H_{u}$ for every $u \in P$ and $a \in G$, where $R_{a}^{*}$ is the tangent mapping to the point transformation $R_{a}u=R_{a}(u)=ua$ of the manifold $P$ induced by $a \in G$;
\item an assignment of the subspace $H_{u}$ is a smooth mapping i.e. a distribution $u \rightarrow H_{u}$ is differentiable.
\end{enumerate}
\end{definicja}

The second condition means that the distribution $u \rightarrow H_{u}$ is invariant by $G$. \\
In the above definition $V_{u}$ denotes a vertical subspace of the space $T_{u}P$, and $H_{u}$ is a horizontal subspace. 

A vector $X \in T_{u}P$ is called horizontal if it is an element of the subspace $H_{u}P$ or vertical if it is an element of the subspace $V_{u}P$. Every vector from the space $T_{u}P$ can be uniquely written in the following way
$$X=Y+Z \ ,$$
where $Y$ is a vertical component and $Z$ a horizontal component of the vector $X$ and they are denoted respectively $vX$ i $hX$.
The third condition in the above definition means simply that if a vector field $X$ is differentiable on $P$ then $vX$ i $hX$ are also differentiable.  \\
\newpage
\begin{figure}[h]
\centering
\includegraphics[width=0.5\textwidth]{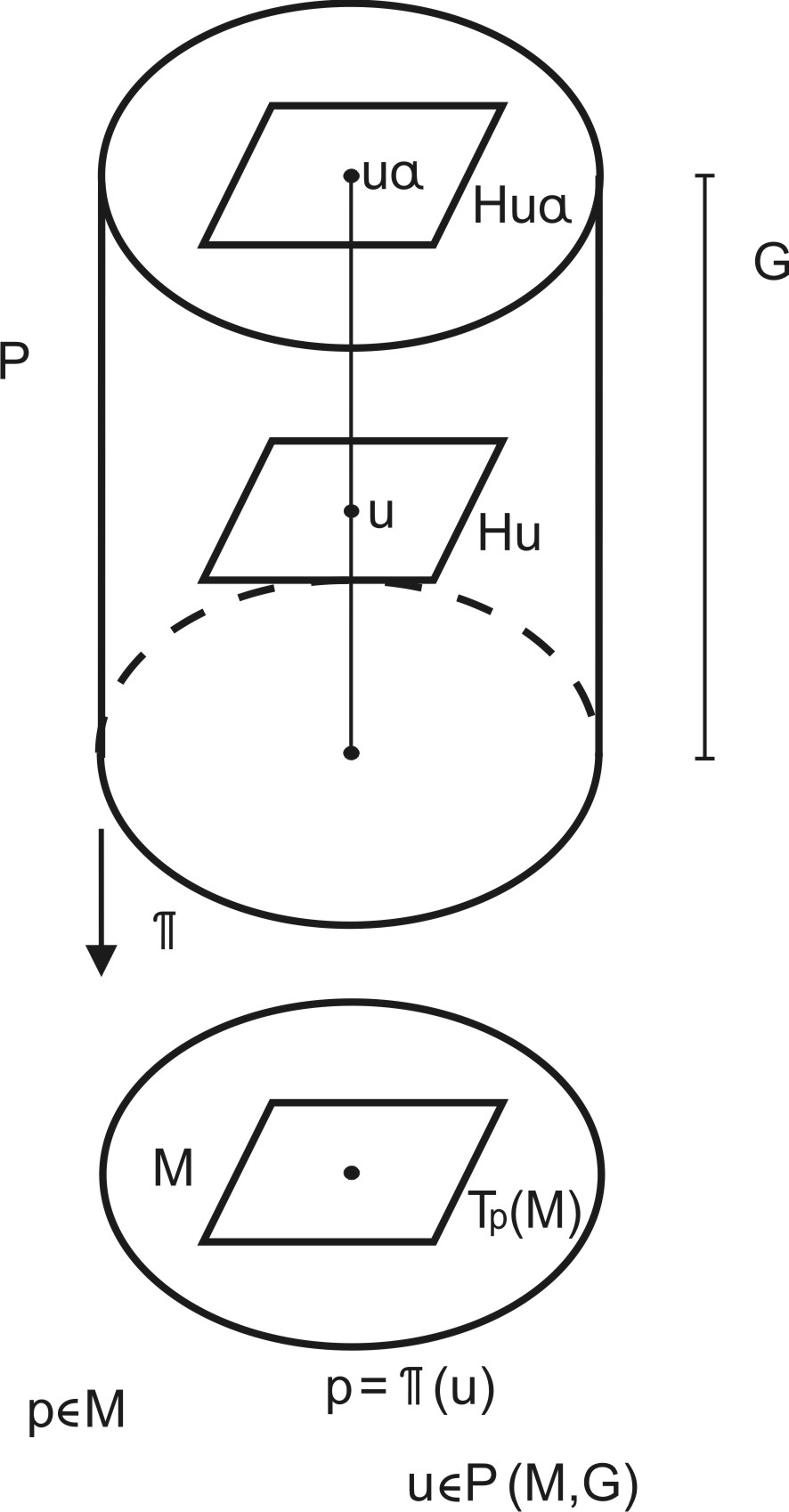}
\caption{Picture ilustrating Definition 13}
\end{figure}

Given a connection $\Gamma $ in $P$ we define a 1-form $\omega $ on $P$ with values in the Lie algebra $\mathfrak{g}$ of the group $G$ as follows. In the section 1.2 we showed that every vector field $A \in \mathfrak{g}$ induces a vector field $A^{* }$ on $P$ which is called the fundamental vector field corresponding to $A$, and the mapping $A \rightarrow (A^{* })_{u}$ is a linear isomorphism from the algebra $\mathfrak{g}$ onto $V_{u}$ for every $u \in P$. For each $X \in T_{u}P$ we define $\omega (X)$ to be the unique $A \in \mathfrak{g}$ such that $(A^{* })_{u}$ is equal to the vertical component of the vector X. Notice, that $\omega (X)=0$ if and only if the vector $X$ is horizontal. The form $\omega $ is called a connection form of the given connection $\Gamma $.

\begin{twierdzenie}
The connection form $\omega $ of the connection $\Gamma $ satisfies the following conditions:
\begin{enumerate}
\item $\omega (A^{* })=A$ for every $A \in \mathfrak{g}$;
\item $(R_{a})_{* }\omega =Ad(a^{-1})\omega $ i.e. $\omega ((R_{a})^{* }X)=Ad(a^{-1})\omega (X)$ for each $a \in G$ and for each vector field $X$ on $P$, where $Ad$ denotes the adjoint representation of $G$ in $\mathfrak{g}$.
\end{enumerate}
Conversely, given a 1-form $\omega $ on $P$ with values in $\mathfrak{g}$ and satisfying conditions 1-2, then there exists a unique connection $\Gamma $ in $P$ whose connection form is $\omega $. 
\end{twierdzenie}

$(R_{a})_{*}\omega $ denotes here a pull-back of the form $\omega $ from the point $u\cdot a \in P$ tu the point $u \in P$. \\

\underline{Proof:}\\
\begin{enumerate}
\item Let $\omega $ be a form of some connection $\Gamma $ on a bundle $P(M,G)$. Then (1.) follows directly from the definition of a connection form.
\item Since every vector field $X$ on $P(M,G)$ is the sum $$X=vX+hX,$$ it is sufficient to verify (2.) in two special cases: 
\begin{itemize}
\item $X$ is a horizontal vector field;
\item $X$ is a vertical vector field.
\end{itemize}
\end{enumerate}
Let $X$ be a horizontal vector field. Then the field $(R_{a})^{*}X$ is also horizontal for every $a\in G$ (it follows from (2.) of the definition of a connenction $\Gamma $ ). Hence $\omega [(R_{a})^{*}X]\equiv 0$ and $Ad(a^{-1})[\omega (X)]\equiv 0$. \\
Now, let $X$ be a vertical field. We can assume that $X$ is a fundamental vector field of $A^{*}$. Then $R_{a}^{*}(A^{*})$ is a fundamental vector field corresponding to $Ad(a^{-1})A$. Therefore we have 
$$[(R_{a})_{*}\omega ]_{u}(X)=\omega _{ua}[R^{*}_{a}(X)]=\omega _{ua}(Ad(a^{-1})A)=Ad(a^{-1})A=Ad(a^{-1})[\omega _{u}(X)].$$
Conversely, let $\omega $ be the 1-form on $P(M,G)$ with properties (1.) and (2.). We define $H_{u}:=\lbrace X\in T_{u}(P):\omega (X)=0\rbrace $. The distribution $u\rightarrow H_{u}$ defines the connection $\Gamma $ on $P(M,G)$ and $\omega $ is its form.

The projection $\pi: \ P\rightarrow M$ induces a linear mapping (tangent, a differential $\pi $), which we denoted by $\pi ^{*}:$\\
$\pi ^{*}:\ T_{u}(P)\rightarrow T_{x}M$ for each $u\in P$, $x=\pi (u)$.\\
When on the principal bundle $P(M,G)$ is given a connection $\Gamma $, the differential $\pi ^{*}$ maps isomorphically a horizontal subspace $H_{u}$ onto $T_{x}(M)$, $x=\pi (u)$.\\

The horizontal lift (or just : lift) of a vector field $X$ from $M$ onto $P(M,G)$ is a unique vector field $X^{*}$ on $P(M,G)$: $X^{*}$ is horizontal and $\pi ^{*}(X^{*}_{u})=X_{\pi (u)}$ for every $u\in P$.

\begin{fakt}
Let $\Gamma $ be the connection in the principal fibre bundle $P$ and let $X$ be a vector field which is defined on the manifold $M$. Then there is uniquely defined a horizontal lift $X^{* }$ of the vector field $X$. The lift $X^{* }$ is invariant with respect to $R_{a}^{*}$ for each $a \in G$ and $\pi ^{*}(X^{*}_{u})=X_{\pi (u)}$. Conversely, every horizontal vector field $X^{* }$ on $P$ which is invariant with respect to $G$ is a lift of a vector field $X$ on the manifold $M$.
\end{fakt}

\begin{figure}[h]
\centering
\includegraphics[width=0.5\textwidth]{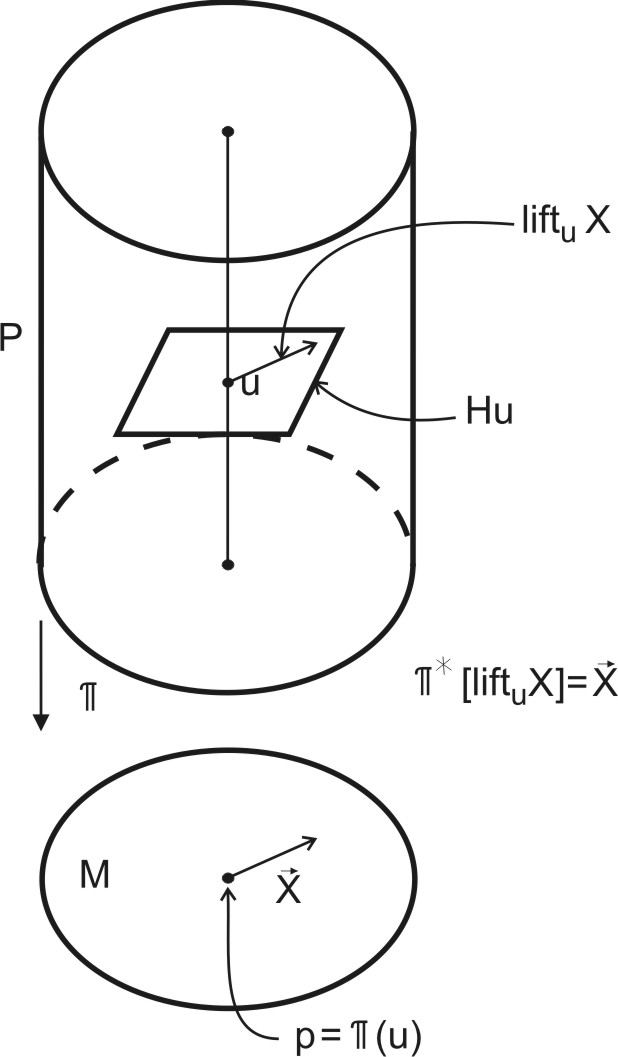}
\caption{Picture ilustrating Fact 2}
\end{figure}

\begin{fakt}
Let $X^{* }$ and $Y^{* }$ be horizontal lifts of the vector fields  $X$ and $Y$ respectively. Then:
\begin{enumerate}
\item $X^{* }+Y^{* }$ is a horizontal lift of the vector field $X+Y$;
\item For every function $f$ on the manifold $M$ $f_{* }\cdot X^{* }$ is a horizontal lift of the vector field $fX$, where $f_{* }$ is the function on $P$ defined by the formula ($f_{*}(u):=f\cdot \pi (u)$); 
\item the horizontal component of the vector field $[X^{* }, Y^{*}]$ is a horizontal lift for $[X,Y]$.
\end{enumerate}
\end{fakt}

Let $(x^{1},...,x^{n})$ be a local coordinate system in a coordinate neighborhood $U\subset M$ and let $X^{*}_{i}\ (i=1,...,n)$ be the horizontal lifts to $\pi ^{-1}(U)$ for the vector fields $X_{i}=\frac{\partial }{\partial x^{i}}$ on $U$ $(i=1,...,n)$. Then the vector fields $X^{*}_{1},...,X^{*}_{n}$ form a basis for the distribution $u\rightarrow H_{u}$ in $\pi ^{-1}(U)$.

Let $c:[0,1]\rightarrow M$  be a curve on $M$ and let $e\in \pi ^{-1}[c(0)]$.

\begin{definicja}
A horizontal lift (a lift) $\bar{c}:\ [0,1]\rightarrow P$ of the curve $c:[0,1]\rightarrow M$ on $P(M,G)$ is a curve $\bar{c}$ on $P(M,G)$ with following properties:\\
\begin{itemize}
\item $\bar{c}(0)=e$;
\item $\pi \cdot \bar{c}=c$;
\item $\bar{c}$ is a horizontal curve, i.e. a vector tangent to $\bar{c}$ at the point $\bar{c}(t)$ belongs to $H_{\bar{c}(t)}$ for every $t\in [0,1].$
\end{itemize}
\end{definicja}

For any $e\in P(M,G)$: $\pi (e)=c(0)$ there is a unique horizontal lift $\bar{c}=c(t)$ of the curve $c:[0,1]\rightarrow M$, which "begins" at the point $e$.
\begin{uwaga}
A horizontal lift of the curve is strictly associated with a lift of the vector field. Namely, if $X^{*}$ is a lift of the vector field $X$ on $M$, then the integral curve of the field $X^{*}$ through $u_{0}\in P(M,G)$ is a lift of the integral curve of the field $X$ through the point $x_{0}=\pi (u_{0})\in M.$
\end{uwaga}

\begin{definicja}
We say that $\bar{c}(1)$ is given by the parallel transport of $\bar{c}(0)=e$ along the curve $c:[0,1]\rightarrow M$.
\end{definicja}

\begin{uwaga}
In general, even if $c: [0,1]\rightarrow M$ is closed (a loop), $c(0)=c(1)$, then $\bar{c}(t)$ is not closed: $\bar{c}(1)=\bar{c}(0)\cdot a$, where $a\in G$. 
\end{uwaga}
A holonomy group $\phi (x_{0})$ of the connection $\Gamma $ at the point $c(0)=x(0)=:x_{0}$ consists of elements of the structure group $G$, which are given by $\bar{c}(1)=\bar{c}(0)\cdot a$ for all loops $c:[0,1]\rightarrow M:\ c(0)=c(1)$, which start and end at the point $x(0)=c(0)=:x_{0}$.

Let us express the form of the connection $\omega $ on $P(M,G)$ by the family of forms defined on open sets of the base differentiable manifold $M$.\\
Let $\lbrace U_{\alpha }\rbrace $ be an open covering $M$ with the family of difeomorphisms $\psi _{\alpha }:\ \pi ^{-1}(U_{\alpha })\rightarrow U_{\alpha }\times G$ and corresponding to this family the family of transition functions $\psi _{\beta \alpha }[\pi (u)]:=\varphi _{\beta }(u)\cdot [\varphi _{\alpha }(u)]^{-1}$, $\varphi _{\alpha }:\ \pi ^{-1}(U_{\alpha })\rightarrow G$, $\psi _{\alpha \beta }: U_{\alpha }\bigcap U_{\beta }\rightarrow G$. For every $\alpha $ let $\sigma _{\alpha }:\ U_{\alpha }\rightarrow P$ be a section of the bundle $P$ over $U_{\alpha }$, defined as: $\sigma _{\alpha }(x):=\psi ^{-1}_{\alpha }(x,e)$; $x\in U_{\alpha }$, and let $e$ be the unit element of the group $G$.\\
$\psi _{\alpha }:\ \pi ^{-1}(U_{\alpha })\rightarrow U_{\alpha }\times G$ $\ \ \rightarrow \ \ $ $\psi ^{-1}_{\alpha }(x,e):\ U_{\alpha }\times \lbrace e\rbrace \equiv U_{\alpha }\rightarrow \pi ^{-1}(U_{\alpha })\subset P$.\\
Let $\theta $ denote a left-invariant and $\frak{g}$-valued cannonical 1-form on the group $G$ [this form is uniquely defined by $\theta (A)=A$ for $A\in \frak{g}$].\\
For every nonempty $U_{\alpha }\bigcap U_{\beta }$ we define $\frak{g}$-valued 1-form $\theta _{\alpha \beta }:$ $\ \ \theta _{\alpha \beta }:\ T(U_{\alpha }\bigcap U_{\beta })\rightarrow \frak{g}$, $\theta _{\alpha \beta }:=(\psi _{\alpha \beta })_{*}\theta $ ($\leftarrow $ a pull-back of the form $\theta $ on $M$), and on every $U_{\alpha }$ we define $\frak{g}$-valued 1-form $\omega _{\alpha }=(\sigma _{\alpha })_{*}\omega $.

\begin{fakt}
Local forms on $M$, $\theta _{\alpha \beta }$ and $\omega _{\alpha }$, satisfy on  $U_{\alpha }\cap U_{\beta }$ conditions:
$$\omega _{\beta }(X)=Ad (\psi_{\alpha \beta }(X))^{-1}[\omega _{\alpha }(X)]+\theta _{\alpha \beta }(X)$$
for each $X \in T_{x}(U_{\alpha}\cap U_{\beta})$, $x \in U_{\alpha}\cap U_{\beta}$.

Conversely, every family of 1-forms $\lbrace \omega_{\alpha }\rbrace $ with values in $\mathfrak{g}$ ($\omega_{\alpha}$ defined on $U_{\alpha}$), which satisfies the above conditions determines a unique 1-form of the connection $\omega $ on $P(M, G)$.
\end{fakt}

 This form generates a family {$\omega_{\alpha}$} in the way given above.

\begin{center}
\underline{Curvature form and structure equations.}
\end{center}
Let $P(M,G)$ be a principal fibre bundle and let $\rho $ be a representation of the structure group $G$ of this bundle in a finite dimensional vector space $V$: $\rho : G\rightarrow GL(V)$; $\rho (a):V\rightarrow V$ is a linear mapping of $V$ for each $a \in G$ such that $\rho (a\cdot b)=\rho (a)\cdot \rho (b)$, $\rho (a)=A\in GL(V)$, $\rho (b)=B \in GL(V)$, $\rho(a\cdot b)=A\cdot B\in GL(V)$. Here $GL(V)$ means the group of linear transformations acting on $V$.

\begin{definicja}
Pseudotensorial form of the type $(\rho ,V)$ and of the degree r on $P(M,G)$ is a $V$-valued r-form $\varphi $ on $P(M,G)$ with property $(R_{a})_{*}\varphi =\rho (a^{-1})\cdot \varphi $, for $a \in G$.
\end{definicja}
In the extended form there is\\
$[(R_{a})_{*}\varphi ]_{u}(X_{1},...,X_{r})=\rho (a^{-1})\cdot \varphi _{u}(X_{1},...,X_{r})$, where $X_{1},...,X_{r}\in T_{u}(P)\times ...\times T_{u}(P)$, $\rho (a^{-1})\in GL(V)$.

\begin{definicja}
A form $\varphi $ of degree r and of the type $(\rho ,V)$ on $P(M,G)$ is tensorial, if it is a horizontal form, i.e. if $\varphi (X_{1},...,X_{r})=0$ if at least one of the tangent vectors  $X_{i}$ ($i=1,...,r$) on $P(M, G)$ is vertical, i.e. tangent to a fibre.
\end{definicja}

Let $\Gamma $ be a connection on $P(M, G)$.  Let $V_{u}$ and $H_{u}$ be a vertical $(V_{u})$ and a horizontal space $(H_{u})$ of the tangent space $T_{u}(P)$ respectively and let $h: T_{u}(P)\rightarrow H_{u}$ be a projection onto $H_{u}$ ($h$ assigns for every $X \in T_{u}(P)$ a horizontal component $hX$ ).

\begin{fakt}
If $\varphi $ is a pseudotensorial r-form of the type $(\rho ,V)$ on $P(M,G)$, then:
\begin{enumerate}
\item The form $\varphi \cdot h$ defined by $\varphi \cdot h(X_{1},...,X_{r}):=\varphi (hX_{1},...,hX_{r})$, $X_{i\ (i=1,...,r)}\in T_{u}(P)$ is a tensorial form of the type $(\rho ,V)$ on $P(M,G)$;
\item $d\varphi $ is a pseudotensorial $(r+1)$-form of the type $(\rho ,V)$;
\item $\mathcal{D}\varphi :=(d\varphi )\cdot h$ is a tensorial $(r+1)$-form of the type $(\rho ,V)$. 
\end{enumerate}
\end{fakt}

\begin{definicja}
A form $\mathcal{D}\varphi :=(d\varphi )\cdot h=d\varphi (hX_{1},...,hX_{r},hX_{r+1})$ is called an exterior covariant differential of an r-form $\varphi $ and the operation $\mathcal{D}$ is called an exterior covariant differentiation.
\end{definicja}

If $\rho $ is an adjoint representation of the group G in its algebra $\mathfrak{g}$, then pseudotensorial form of the type $(\rho ,\mathfrak{g})$ is called a form of the type $AdG$.\\
An example: a connection form $\omega $ is a pseudotensorial 1-form of the type $AdG$.

\begin{fakt}
$\mathcal{D}\omega =(d\omega )h=d\omega (hX)$ is a tensorial 2-form of the type $AdG$ which is called a curvature form of the connection $\omega $. We denote it by 2-form $\Omega $.
\end{fakt}

\begin{definicja}
Let $\omega $ be a connection form and $\Omega $ its curvature form. Then an equation $\Omega (X,Y)=d\omega (X,Y)+\frac{1}{2}[\omega(X),\omega(Y)]$ is a structure equation of the connection $\omega $, $X, Y \in T_{u}(P)$ and $u\in P(M,G)$. 
\end{definicja}
\begin{twierdzenie}
Let $\vec{e}_{1},...,\vec{e}_{r}$ be a basis of a Lie algebra $\mathfrak{g}$ of a Lie group $G$, and let $C^{i}_{jk}=-C^{i}_{kj}$ $(i, j,k=1,...,r)$ be structure constants of the algebra $\mathfrak{g}$ with respect to this basis, i.e. $[\vec{e}_{j},\vec{e}_{k}]=C^{i}_{jk}\vec{e}_{i}$, $(i,j,k=1,...,r)$. Let $\omega =\omega ^{i}\vec{e}_{i}$, $\Omega =\Omega ^{j}\vec{e}_{j}$ for $i,j=1,...,r$. Then the structure equation of the connection $\omega $ can be expressed as follows:
$$d\omega ^{i}=-\frac{1}{2}C^{i}_{jk}\omega^{j}\wedge\omega^{k}+\Omega^{i} \ \ (i=1,...,r).$$
\end{twierdzenie}
\newpage
\underline{Proof:}\\
\begin{align}
&(d\omega ^{i})\vec{e}_{i}=-\frac{1}{2}[\omega ^{i}\vec{e}_{i},\omega ^{k}\vec{e}_{k}]+\Omega ^{i}\vec{e}_{i} \notag
\\&[\omega ^{i}\vec{e}_{i},\omega ^{k}\vec{e}_{k}]=[\vec{e}_{i},\vec{e}_{k}]\omega ^{i}\wedge\omega ^{k} \notag
\\&(d\omega ^{i})\vec{e}_{i}=-\frac{1}{2}[\vec{e}_{i},\vec{e}_{k}]\omega ^{i}\wedge\omega ^{k}+\Omega ^{i}\vec{e}_{i}\notag
\\&[\vec{e}_{i},\vec{e}_{k}]=C^{l}_{ik}\vec{e}_{l} \notag
\\&(d\omega ^{i})\vec{e}_{i}=-\frac{1}{2}C^{t}_{lk}\omega ^{l}\wedge\omega ^{k}\vec{e}_{t}+\Omega ^{i}\vec{e}_{i}\notag
\\&d\omega ^{i}\vec{e}_{i}=-\frac{1}{2}C^{i}_{lk}\omega ^{l}\wedge\omega ^{k}\vec{e}_{i}+\Omega ^{i}\vec{e}_{i}\notag
\\&d\omega ^{i}\vec{e}_{i}=(-\frac{1}{2}C^{i}_{lk}\omega ^{l}\wedge\omega ^{k}+\Omega ^{i})\vec{e}_{i},\notag
\end{align}

or after missing basis $\lbrace e_{i}\rbrace $

\begin{align}
d\omega ^{i}=-\frac{1}{2}C^{i}_{lk}\omega ^{l}\wedge\omega ^{k}+\Omega ^{i}\notag
\end{align}

We have an important Bianchi identity:$\mathcal{D}\Omega \equiv 0$.\\
\underline{Proof:}\\
\begin{align}
\mathcal{D}\Omega (X,Y,Z):=d\Omega \cdot h(X,Y,Z)=d\Omega \cdot h=d\Omega (hX,hY,hZ). \notag 
\end{align}
To prove Bianchi identity it is sufficient  to show that $d\Omega (X,Y,Z)\equiv 0$, where $X$, $Y$, $Z$ are horizontal vector fields. We apply the external differentiation $d$ to structure equations $d\omega ^{i}=-\frac{1}{2}C^{i}_{lk}\omega ^{l}\wedge \omega ^{k}+\Omega ^{i}$, and then we obtain $d\cdot d\omega ^{i}\equiv 0=-\frac{1}{2}C^{i}_{lk}(d\omega ^{l}\wedge \omega ^{k}-\omega ^{l}\wedge d\omega ^{k})+d\Omega ^{i}$. Since $\omega ^{i}(X)=0$ for each horizontal field $X$, then also $(d\omega ^{l}\wedge \omega ^{k}-\omega ^{l}\wedge d\omega ^{k})(X,Y,Z)\equiv 0$ because we in here have expressions $\omega ^{k}(Z)$, $\omega ^{l}(X)$. Thus we obtain\\
$d\Omega ^{i}(hX,hY,hZ)=0 \implies d\Omega (hX,hY,hZ)=\mathcal{D}\Omega (X,Y,Z)\equiv 0$\\

\begin{center}\subsection{Linear connection.} \end{center}

Now we are going to consider a connection in the bundle of linear frames $L(M)$.  Take $P$ which will denote $L(M)$ in our further considerations, let denote by $G$ the  general linear group $GL(n, \mathbb{R})$, where $n=dimM$. \\
\begin{definicja}
A canonical form $\theta $ on $P$ is a 1-form on $P$ with values in $\mathbb{R}^{n}$ and it is defined as follows
$$\theta (X)=u^{-1}(\pi^{*}(X))\ \ \ for\ X \in T_{u}(P), $$
where $u$ is a linear mapping  $u: \mathbb{R}^{n}\rightarrow T_{\pi (u)}(M)$, $x=\pi (u)$, $\pi ^{*} (X) \in T_{x}(M)$.
\end{definicja}

$\theta $ is also called a soldering (or a solder form). It joins the bundle of linear frames $P[\equiv L(M)]$ with the base manifold $M$ and causes that the geometrical structure of the base is determined by the geometrical structure of the bundle (and conversely). It follows from the fact that $M$ is modelled on $\mathbb{R}^{n}$, and $\theta $ has values in $\mathbb{R}^{n}$. Other general principal fibre bundles over $M$ do not have such a form. In a natural basis $\{\vec{e}_{i}\}_{(i=1,...,n)}$ of the space $\mathbb{R}^{n}$ $\theta =\theta ^{i}\vec{e}_{i}$, where $\theta ^{i}_{\ \ (i=1,...,n)}$ is a set of $n$ 1-forms with values in $\mathbb{R}$. If $X\in T_{u}[L(M)]$, then $\theta ^{i}(X)$ is the i-th component of the projection of $X$ onto $M$ in the basis $u=(X_{1},...,X_{n})$ of the space $T_{x}(M)$: $X_{i}=u(e_{i})$; $x=\pi (u)$.

\begin{fakt}
A canonical form $\theta $ in $P[\equiv L(M)]$ is a tensorial 1-form (horizontal) of the type $[id, \mathbb{R}^{n}]$. 
\end{fakt}

\underline {Proof:}\\
Let $X$ be a vertical vector at $u\in P$. Then $\pi^{*}(X)=0$; $\theta (X)=u^{-1}[\pi ^{*}(X)]=u^{-1}(0)=0$. Hence $\theta $ is a tensorial 1-form (horizontal). If $X\in T_{u}(P)$ is any vector at $u\in P$ and $a\in G$ is any element of $G\equiv GL(n, \mathbb{R})$, then $R_{a}^{*}(X)$ is a vector at $u\cdot a \in P$. Therefore $[(R_{a})_{*}\theta ](X)=\theta [R^{*}_{a}(X)]=(ua)^{-1}[\pi ^{*}(R^{*}_{a}(X))]=a^{-1}\cdot u^{-1}[\pi ^{*}(X)]=a^{-1}\cdot \theta (X)=\rho (a^{-1})\theta (X)$. Since $\rho (a^{-1})=a^{-1}$, $\rho =id$. Thus $\theta $ is a tensorial 1-form of the type $(id, \mathbb{R}^{n})$ on $P$.

\begin{definicja}
A connection in the bundle $L(M)$ is called a linear connection of the manifold $M$. 
\end{definicja}
A linear connection $\Gamma $ of the manifold $M$ allows to assign for every $\xi \in \mathbb{R}^{n}$ a horizontal vector field $B(\xi )$ on the bundle $P[\equiv L(M)]$ as follows:\\
for each $u\in L(M)$ $[B(\xi )]_{u}$ there is a unique horizontal vector at $u$ such that $\pi ^{*}([B(\xi )]_{u})=u(\xi )$, $u(\xi )\in T_{x}(M)$, $x=\pi (u)$. 

\begin{definicja}
A vector field $B(\xi )$ is called a standard horizontal vector field on $L(M)$, corresponding to $\xi \in \mathbb{R}^{n}$.
\end{definicja}

\begin{uwaga}
A standard horizontal vector field depends on the choice of a connection in the bundle $L(M)$. (Fundamental fields on $L(M)$ did not depend on the connection.)
\end{uwaga}

\begin{fakt}
A standard horizontal vector field has the following properties:
\begin{enumerate}
\item if $\theta $ is a canonical 1-form on the bundle $L(M)$, then $\theta [B(\xi )]=\xi$ for $\xi \in \mathbb{R}^{n}$;
\item $R_{a}^{*}[B(\xi )]=B(a^{-1}\xi )$ for $a \in G$ and $\xi \in \mathbb{R}^{n}$, $a^{-1}(\xi )\in \mathbb{R}^{n}$, $G=GL(n,\mathbb{R})$;
\item if $\xi \neq 0$, then $B(\xi )$ is everywhere $\neq 0$.
\end{enumerate}
\end{fakt}

\begin{uwaga}
Conditions $\theta (B(\xi ))=\xi $ and $\omega (B(\xi ))=0$ (where $\omega $ is a connection form) determine $B(\xi )$ for each $\xi \in \mathbb{R}^{n}$.
\end{uwaga}

Let $(B_{1},...,B_{n})$ be standard horizontal vector fields corresponding to the natural basis $\vec{e}_{1},...,\vec{e}_{n}$ in $ \mathbb{R}^{n}$ and let $\lbrace E^{\ j*}_{i}\rbrace $ be fundamental vector fields, which correspond to the basis $\lbrace E^{\ j}_{i}\rbrace $ in $gl(n, \mathbb{R})$. Then $\lbrace B_{i},E^{\ j*}_{i}\rbrace $ and $\lbrace \theta ^{i},\omega ^{i}_{\ j}\rbrace $ are dual to each other in the following sense
$$\theta ^{k}(B_{i})=\delta ^{k}_{i}, \ \ \theta ^{k}(E^{\ j*}_{i})=0,$$
$$\omega ^{\ k}_{l}(B_{i})=0, \ \ \omega ^{\ k}_{l}(E^{\ j*}_{i})=\delta ^{k}_{i}\delta ^{j}_{l}.$$
\begin{uwaga}
$n^{2}+n$ many vector fields $\lbrace B_{k},\ E_{i}^{\ j*}, \ i,j,k=1,...,n \rbrace $ introduce in $L(M)$ teleparalelism, i.e. that $n^{2}+n$ many vectors $[(B_{k})_{u},(E_{i}^{\ j*})_{u}]$ form a basis of $T_{u}(P)$ for each $u\in P\equiv L(M)$. It follows from the above remark, that the bundle $T[L(M)]$, which is tangent to $L(M)$, is trivial.
\end{uwaga}

\begin{definicja}
A 2-form $\Theta :=\mathcal{D}\theta $ is called a 2-form of torsion of the linear connection $\omega $ on $L(M)$. $\Theta $ is a tensorial 2-form on $L(M)$ of the type $(id, \mathbb{R}^{n})$.
\end{definicja}

\begin{fakt}
If $A^{\ast }$ is a fundamental vector field corresponding to $A \in \mathfrak{g}$ and if $B(\xi )$ is a standard vector field corresponding to $\xi \in \mathbb{R}^{n}$, then
$$[A^{\ast }, B(\xi )]=B(A\xi )\ ,$$
where $A\xi $ indicates the image of $\xi $ with respect to $A \in \mathfrak{g}=\mathfrak{gl}(n, \mathbb{R})$ (a Lie algebra of all matrices of the dimension $n \times n$ ), which acts in $\mathbb{R}^{n}$. 
\end{fakt}

\begin{fakt}
(Structure equations of the linear connection) Let $\omega ,\ \Theta $ and $\Omega $ be a connection form, a torsion form and a curvature form of the linear connection $\Gamma $ on the manifold $M$ respectively. We have \\
the first structure equation:\\
$$\Theta (X, Y)\ =d\theta (X, Y)+\frac{1}{2}\Bigl (\omega(X)\cdot \theta (Y) - \omega (Y)\cdot \theta (X)\Bigr) ,$$
and the second structure equation:\\
$$\Omega (X,Y)=d\omega (X, Y)+\frac{1}{2}\Bigl [\omega (X), \omega (Y)\Bigr ],$$
where $X,\ Y \in T_{,}(P)$ and $u \in P[\equiv L(M)]$.
\end{fakt}

In the basis of $\mathbb{R}^{n}$ and $\mathfrak{gl}(n, \mathbb{R})$ we have $\Theta ^{i}=d\theta ^{i}+\omega ^{i}_{.j}\wedge \theta ^{j}$, $\Omega ^{i}_{\ j}=d\omega ^{i}_{.j}+\omega ^{i}_{.k}\wedge \omega ^{k}_{.j}$.

\underline{Proof:}\\
$$\Theta (X, Y)\ =d\theta (X, Y)+\frac{1}{2}\Bigl (\omega(X)\cdot \theta (Y) - \omega (Y)\cdot \theta (X)\Bigr) ,$$
$X, Y\in T_{u}(P)$, $P\equiv L(M)$ - a principal bundle of linear frames over $M$, $\theta (X) \in \mathbb{R}^{n}$, $\omega (X) \in \mathfrak{gl}(n, \mathbb{R})$. \\
In the basis of $\mathbb{R}^{n}$ and $\mathfrak{gl}(n, \mathbb{R})$ we have:\\
$\theta =\theta ^{i}\vec{e}_{i}$, $\Theta =\theta ^{i}\vec{e}_{i}$, $\omega =\omega ^{i}_{.j}E^{\ j}_{i}$, $\Omega =\Omega ^{i}_{.j}E^{\ j}_{i}$ for $(i,j=1,...,n)$.
$$\Theta ^{i}(X, Y)\vec{e}_{i} =d\theta ^{i}(X, Y)\vec{e}_{i}+\frac{1}{2}\Bigl (\omega^{i}_{.l}(X)\cdot \theta ^{l}(Y) - \omega ^{i}_{.l} (Y)\wedge \theta ^{l}(X)\Bigr)\vec{e}_{i} ,$$
$$\Theta ^{i}(X, Y)\vec{e}_{i} =d\theta ^{i}(X, Y)\vec{e}_{i}+\omega^{i}_{.l}(X)\wedge \theta ^{l}(Y)\vec{e}_{i} .$$
Hence $\Theta ^{i}=d\theta ^{i}+\omega^{i}_{.l}\wedge \theta ^{l}.$\\
$\theta $ - a canonical 1-form in the principal bundle $L(M)$, $\Theta =\mathcal{D}\theta $ - 2-form of torsion of the linear connection.
$$\Omega (X,Y)=d\omega (X, Y)+\frac{1}{2}\Bigl [\omega (X), \omega (Y)\Bigr ],\ \omega =\omega ^{ij}E_{ij}=\omega ^{i}_{\ j}E^{\ j}_{i},$$
\begin{align}
\Omega ^{i}_{.j}(X,Y)E^{\ j}_{ i}&=d\omega ^{i}_{.j}(X, Y)E^{\ j}_{i}+\frac{1}{2}\Bigl [\omega ^{i}_{.j}(X)E^{\ j}_{i}, \omega ^{k}_{.l}(Y)E^{\ l}_{k}\Bigr ] \notag
\\&=d\omega ^{i}_{.j}(X, Y)E^{\ j}_{i}+\frac{1}{2}\Bigl [E^{\ j}_{i}, E^{\ l}_{k}\Bigr ]\omega ^{i}_{.j}\wedge \omega ^{k}_{.l} (X,Y).\notag
\end{align}
We made use of the fact:\\
\begin{align}
\Bigl [\omega (X), \omega (Y)\Bigr ]:&=\Bigl [\omega ^{l}_{\ j}(X)E^{\ j}_{i}, \omega ^{k}_{\ l}(X)E^{\ l}_{k}\Bigr ] \notag
\\&=\Bigl [E^{\ j}_{i}, E^{\ l}_{k}\Bigr ]\omega ^{i}_{\ j}\wedge  \omega ^{k}_{\ l}(X,Y), \notag
\end{align}
$X,Y \in T_{u}(L(M))$. Then we use the fact, that the commutator\\
$\Bigl [E^{\ j}_{i}, E^{\ l}_{k}\Bigr ]=\delta ^{j}_{k}E^{\ l}_{i}-\delta ^{l}_{i}E^{\ j}_{k}$,\\
and obtain:
\begin{align}
\Omega ^{i}_{.j}(X,Y)E^{\ j}_{i}&=d\omega ^{i}_{.j}(X, Y)E^{\ j}_{i}+\frac{1}{2}\Bigl (\delta ^{j}_{k}E^{\ l}_{i}-\delta ^{l}_{i}E^{\ j}_{k}\Bigr )\omega ^{i}_{.j}\wedge \omega ^{k}_{.l}(X,Y) \notag
\\&=d\omega ^{i}_{.j}(X, Y)E^{\ j}_{i}+\frac{1}{2}\Bigl (\omega ^{i}_{.k}\wedge \omega ^{k}_{.j}(X,Y)E^{\ j}_{i}-\omega ^{l}_{.j}\wedge \omega ^{i}_{.l}(X,Y)E^{\ j}_{i}\Bigr ) \notag
\\&=d\omega ^{i}_{.j}(X, Y)E^{\ j}_{i}+\frac{1}{2}\Bigl (\omega ^{i}_{.k}\wedge \omega ^{k}_{.j}-\omega ^{k}_{.j}\wedge \omega ^{i}_{.k}\Bigr ) (X,Y)E^{\ j}_{i}\notag
\\&=d\omega ^{i}_{.j}(X, Y)E^{\ j}_{i}+\omega ^{i}_{.k}\wedge \omega ^{k}_{.j}(X,Y)E^{\ j}_{i}. \notag
\end{align}
From there after missing the basis $\lbrace E^{\ j}_{i}\rbrace $ we get
$$\Omega ^{i}_{.j}=d\omega ^{i}_{.j}+\omega ^{i}_{.k}\wedge \omega ^{k}_{.j}.$$

The expressions $\Theta ^{i}=d\theta ^{i}+\omega^{i}_{.l}\wedge \theta ^{l}$ and $\Omega ^{i}_{.j}=d\omega ^{i}_{.j}+\omega ^{i}_{.k}\wedge \omega ^{k}_{.j}$ are used to practical computations of the curvature form or the torsion form of the linear connection.

\begin{twierdzenie}
(Bianchi identity) For the linear connection we have:\\
\underline{the first identity of Bianchi}\\
$$3D\Theta (X,Y,Z)\equiv \Omega (X,Y)\theta (Z)+\Omega (Y,Z)\theta (X)+\Omega (Z,X)\theta (Y)\ $$
or equivalently in the terms of basis of the algebra's $gl(n,\mathbb{R})$ and $\mathbb{R}^{n}$
$$D\Theta ^{i}\equiv \Omega ^{i}_{k}\wedge \theta ^{k},$$
where $X,\ Y,\ Z \in T_{u}(P)$, $P\equiv L(M)$.\\
$\theta =1- $ a canonical form on the bundle $L(M).$\\
\underline{the second identity of Bianchi}\\
$$D\Omega \equiv 0\ $$
or in the algebra basis $gl(n, \mathbb{R})$, $\mathcal{D}\Omega ^{i}_{\ k}=0.$
\end{twierdzenie}

\begin{uwaga}
Since $\Omega ^{i}_{.j}$, $\Theta ^{i}$, $\theta ^{i}$ are horizontal, we can decompose $\Omega ^{i}_{.j}$, $\Theta ^{i}$ into components in the basis $\theta ^{i}\wedge \theta ^{l}$, $(i,l=1,...,n)$.
$$\Omega ^{i}_{.j}=\frac{1}{2}R^{i}_{.jkl}\theta ^{k}\wedge \theta ^{l},$$
$$\Theta ^{i}=\frac{1}{2}Q^{i}_{.kl}\theta ^{k}\wedge \theta ^{l}.$$
\end{uwaga}
The above decompositions define the following tensors: curvature tensor $R^{i}_{.jkl}=-R^{i}_{.jlk}$ and torsion tensor $Q^{i}_{.kl}=-Q^{i}_{.lk}$ of the linear connection [on the bundle $L(M)$].

\begin{center}\subsection{Metric connection.} \end{center}

In the previous section we defined a linear connection. A linear connection of a manifold $M$ defines for every curve $\tau =x_{t},\ 0 \le t \le 1$ a parallel displacement of a tangent space $T_{x_{0}}(M)$ onto a tangent space $T_{x_{1}}(M)$. The tangent spaces are considered here as a vector spaces and the parallel displacement is a linear isomorphism between them.

\begin{definicja}
A metric or a metric tensor of the class $C^{k}$ on a differential manifold ($M_{n},A_{M}$) is a tensorial field $g$ of the class  $C^{k}$ and the type ($0,2$) satisfying conditions [1], [2], [4]:
\begin{enumerate}
\item $g$ is symmetric, i.e.\ $ \forall \  x \in M_{n}$ a tensor $g_{x}$ ($g_{x}=g$ at the point $x$) is a symmetric tensor $g_{x}(u,v)=g_{x}(v,u), u,v \in T_{x}(M_{n}), x \in M_{n}$
\item $\forall \ x \in M_{n}$ a bilinear form $g_{x}$ is nondegenerate, i.e. $g_{x}(u,v)=0$, $ v,\ u \in T_{x}(M_{n}), \forall \ v \in T_{x}(M_{n})$ iff $u=0$. The form $g_{x}(u,v)$ defines a scalar product in $T_{x}(M_{n})$. In the terms of the components of the vectors $u$ and $v$ it has a form  $g_{x}(u,v)=g_{ik}u^{i}v^{k}$, where $g_{ik}=g_{x}\left (\partial _{i},\partial _{k}\right )$.

\end{enumerate}
\end{definicja}

\begin{definicja}
A differential manifold with such defined metric is called a Riemannian manifold. It is also said, that $g$ equips $M_{n}$ with a Riemannian structure [8], [11].
\end{definicja}

\begin{definicja}
A Riemannian manifold is called a proper Riemannian manifold if and only if $\forall \  0 \neq v \in T_{x}(M_{n})$ and $x \in M_{n}$ it holds that $g_{x}(v,v) > 0$. In the other case we say about a pseudoriemannian manifold [8], [9], [11].
\end{definicja}

\underline{Equivalent definitions of a metric field $g$ which are more suitable } \\
\underline{ in the theory of a linear connection on the bundle $L(M)$ [11].}\\
\begin{itemize}
\item A section of a bundle of symmetrical tensors of the type $(0,2)$ associated with a principal bundle of linear frames $L[M,GL(n,\mathbb{R})]$;\\   
\item A set of $\frac{n\cdot (n+1)}{2}$ functions $g_{ij}:L(M)\rightarrow \mathbb{R}$: $$g_{ij}(u\cdot a)=A^{k}_{\ i}A^{l}_{\ j}g_{kl}(u),$$ where $g_{kl}(u):=g(\vec{e}_{k},\vec{e}_{l})$ (a scalar product $(\vec{e}_{k},\vec{e}_{l})$), $u=\lbrace \vec{e}_{k}\rbrace =$ a frame at $x=\pi (u)$, $(i,j,k=1,...,n)$, $a=(A^{i}_{j})\in GL(n, \mathbb{R})$.\\
\item A reduction of a principal bundle $L(M)$ to the orthogonal group $\mathcal{O}(n)$, which is a subgroup of $GL(n,\mathbb {R})$. Such reduction gives a bundle of an orthonormal frames $O[M,\mathcal{O}(n)]\subset L(M).$ (It is a subbundle of a principal bundle $L[M,GL(n, \mathbb{R})]$). The bundle $O[M,\mathcal{O}(n)]$ defines $g$: if $u=\lbrace \vec{e}_{i}\rbrace \in O[M,\mathcal{O}(n)]$ is given, then at $x=\pi (u)$ $g=e^{1}\otimes e^{1}+...+e^{n}\otimes e^{n}$, where $e^{i}(\vec{e}_{j})=\delta ^{i}_{j}$ $\bigl [ \lbrace \vec{e}_{i}\rbrace $ is a frame in $T_{x}(M)$; $\lbrace e^{j}\rbrace $ is a dual basis to $\lbrace \vec{e}_{i}\rbrace $, i.e. a basis in $T^{*}_{x}(M) \bigr ]$. Conversely, if there is given a metric $g$ on $M$, then $O[M,\mathcal{O}(n)]$ is defined as a set of all orthonormal frames with respect to $g$:
$$O[M,\mathcal{O}(n)]=\lbrace u\in L(M): g_{ij}(u)=\delta _{ij}\rbrace ,$$
$$u=(e_{1},...,e_{n})=:\lbrace e_{i}\rbrace ,$$
$$g_{ij}(u)=g(\vec{e}_{i},\vec{e}_{j}).$$
$O[M,\mathcal{O}(n)]$ is a set of pairs $(x, \mathcal{O}_{x})$, where $x\in M_{n}$ and $\mathcal{O}_{x}$ is any orthonormal frame at a point x.
\end{itemize}

\begin{uwaga}
If the metric $g$ on $M$ is indefinite, i.e. if the quadratic form $g_{ij}(x)dx^{i}dx^{j}$ is not positive definite and has a sygnature $(k,l)$, then we construct over $M$ a principal bundle of pseudoorthonormal frames $O[M,\mathcal{O}(k,l)]$ with a pseudoorthogonal structure group $\mathcal{O}(k,l)$. $\mathcal{O}(k,l)$ extends a special pseudoorthogonal group $SO(k,l)$ by including reflections.\\
In the physical spacetime $(M_{4},g_{L})$ $(k=1, l=3)$ a pseudoorthogonal group is the Lorentz group $[\mathcal{L}\equiv SO(1,3)]$. Here we consider a bundle of tetrads $O[M_{4},\mathcal{L}]$ over the spacetime and a dual bundle of Lorentzian corepers $P(M_{4},\mathcal{L})$. $\mathcal{L}$ denotes here a Lorentz group which is isomorphic to the group $SO(1,3)$ and $g_{L}$ means the Lorentzian metric on $M_{4}$. A tetrad at the point $x\in (M_{4},g_{L})$ is a basis $\lbrace \vec {e}_{I}\rbrace \in T_{x}(M_{4},g_{L}):g_{L}(e_{I},e_{K})=\eta _{IK}=diag(1,-1,-1,-1)$, and a Lorentzian coreper $\lbrace \vartheta ^{K}\rbrace $ at $x\in M_{4}$ is a set of four 1-forms: $g=\eta_{IK}\vartheta ^{I}\otimes \vartheta ^{K}$, $\vartheta ^{K}(e_{I})=\delta ^{K}_{I}.$
\end{uwaga}

Orthonormal tetrads and Lorentzian corepers are very useful in GR.
Here and in the future we will denote tetrads and cotetrads indices by using big Latin letters.

\begin{definicja}
A linear connection $\Gamma $ on the principal bundle $L(M)$ is called compatible with the metric $g$ or a metric connection $\iff $ for every $e\in O[M,\mathcal{O}(n)]$ $H_{e}\subset T_{e}O[M,\mathcal(n)].$ For such connection $\mathcal{D}g=0.$ $(=\mathcal{D}g_{ij}=0;$ where $ \mathcal{D}g_{ij}=dg_{ij}-\omega _{ij}-\omega _{ji})$.
\end{definicja}

Below we give the other formulation of the compatibility of the connection $\Gamma $ on $L(M)$ with metric $g$. This formulation is better adjusted to the bundle $L(M)$.

\begin{definicja}
$\Gamma $ is compatible with $g$ $\iff $ $\omega $ is $\mathfrak{o}(n)$-valued on $\mathcal{O}[M_{n},O(n)]$. $\mathfrak{o}(n)$ stands for the algebra of the orthogonal group $O(n)$, i.e. $\omega $ when restricted to $\mathcal{O}[M,O(n)]$ takes values in $\mathfrak{o}(n)$.
\end{definicja}

From all metric connections on the bundle $L(M)$ the most important is \underline{Riemannian connection}, which is also called \underline{Levi-Civita} connection. It is a metric connection ($\mathcal{D}g=0;\ \mathcal{D}g_{ij}=dg_{ij}-\omega _{ij}-\omega _{ji}$), whose torsion equals to zero ($\Theta =\mathcal{D}\theta =0$). This connection is unique and it is completely determined by a metric $g$ and its partial derivatives [J. A. Schouten's theory; see, e.g. [4,5]]. In a relativistic theory of gravity we usually restrict to metric connections.
\newpage
Let $P_{1}(M_{1},G_{1})$ and $P_{2}(M_{2},G_{2})$ be principal fibre bundles.
\begin{definicja}
A homomorphism of principal fibre bundles $P_{1}$ and $P_{2}$ is a triple of mappings $(h,k,f)$ such that $h:P_{1}\rightarrow P_{2}$,  $k:G_{1}\rightarrow G_{2}$ and  $f:M_{1}\rightarrow M_{2}$, where $k$ is a homomorphism of Lie groups, such that the following diagram commutes.
\end{definicja} 

$P_{1}\times G_{1}\xrightarrow {\ \ h\times k\ \ } P_{2}\times G_{2}$

$\ \ R_{1}\ \ \downarrow \ \ \ \ \ \ \ \ \ \ \ \ \downarrow \ \ R_{2}\ \ \ \ \ \ \ \ R_{i}:P_{i}\times G_{i}\rightarrow P_{i},\ i=1,2$

$\ \ \ \ \ \ \ \ P_{1}\ \ \ \ \ \ \ \ \ \ P_{2}$

$\ \ \ \pi _{1}\ \ \downarrow \ \ \ \ \ \ \ \ \ \ \ \ \downarrow \ \ \pi _{2}$

$\ \ \ \ \ \ \ M_{1}\xrightarrow {\ \ \ f\ \ \ } M_{2}$

\begin{twierdzenie}
If there is a homomorphism of bundles between $P_{1}(M_{1},G_{1})$ and $P_{2}(M_{2},G_{2})$, then the connection on the bundle $P_{1}(M_{1},G_{1})$ uniquely determines a connection on the bundle $P_{2}(M_{2},G_{2})$.
\end{twierdzenie}

\underline{Structure equations of Riemannian connection on $L(M)$.}\\
$\Omega (X,Y)=d\omega (X,Y)=\frac{1}{2}[\omega (X),\omega (Y)]$ 2-nd structure equation\\
$d\theta (X,Y)+\frac{1}{2}(\omega (X)\cdot \theta (Y))-\omega (Y)\cdot \theta (Y)=0$  1-st structure equation\\
$X,Y \in T_{u}(L(M))$.\\
In basis of $\mathbb{R}^{n}$ and $gl(n, \mathbb{R})$ we obtain:\\
$\Omega ^{i}_{\ j}=d\omega ^{i}_{\ j}+\omega ^{i}_{\ k}\wedge \omega ^{k}_{\ j},$\\
$d\theta ^{i}+\omega ^{i}_{\ k}\wedge \theta ^{k}=0.$

$\theta ^{i}$ is the i-th component in the natural basis of $\mathbb{R}^{n}$ of canonical 1-form $\theta $ on $L(M)$.

\underline{Bianchi identities:}\\
1-st $\Omega (X,Y)\cdot \theta (Z)+\Omega (Y,Z)\cdot \theta (X)+\Omega (Z,X)\cdot \theta (Y)\equiv 0,$\\ 
or in the basis of $gl(n,\mathbb{R})$ and $\mathbb{R}^{4}$\\
$\mathcal{D}\Theta ^{i}=\Omega ^{i}_{k}\wedge \theta ^{k}=0$.\\
2-nd $\mathcal{D}\Omega (X,Y,Z)\equiv 0$ \\
or in the basis of algebra $gl(n,\mathbb{R})$ $\mathcal{D}\Omega ^{i}_{k}=0$.

\begin{center}
\underline{Expressions in the local charts on $M$.}
\end{center}

Let $(M_{n},A_{M})$ be a differential manifold and $(U,\varphi )$ a local map on $M_{n}$ with local coordinates $(x^{1},...,x^{n})$ on $U\subset M_{n}$. Here $A_{M}$ means the maximal atlas on $M_{n}$. Let denote by $\lbrace X_{i}=\frac{\partial }{\partial x^{i}}\rbrace (i=1,...,n)$ vector fields of natural basis ($\equiv $ coordinates basis) on $U$. Every linear frame $u=(X_{1},...,X_{n})$ at the point $x\in U$ can be uniquely given in the form $X_{k}=X^{i}_{\ k}\partial _{i}$, $\lbrace \partial _{k}\rbrace $ is a natural basis at $x\in V$, $det(X^{i}_{j})\neq 0$ $(i,k=1,...,n)$\\
$\lbrace x^{i},X^{j}_{\ k}\rbrace $ are the local coordinates at $\pi ^{-1}(U)\subset L(M)\ \ (i,j,k=1,...,n).$ Let $[Y^{j}_{\ k}]$ be an inverse matrix to the matrix $[X^{j}_{\ k}]:\ \ X^{j}_{\ i}Y^{k}_{\ j}=Y^{j}_{\ i}X^{k}_{\ j}=\delta ^{k}_{i}$ and let $(\vec{e}_{1},...,\vec{e}_{n})$ be a natural basis at $\mathbb{R}^{n}$. Let $\theta =\theta ^{i}\vec{e}_{i}$, where $\theta $ is a canonical 1-form (soldering form) on the principal fibre bundle of linear frames $L(M)$.

\begin{fakt}
Forms $\theta ^{i} =\theta ^{i}\vec{e}_{i}$ are expressed in the terms of local coordinates as follows $\lbrace x^{i}, X^{j}_{k}\rbrace $
$$\theta ^{i}=Y^{i}_{\ j}dx^{j}.$$
In the natural basis $\lbrace X_{i}=\partial _{i}\rbrace $, $Y^{l}_{\ k}=\delta ^{l}_{k}$ and $\theta ^{i}=dx^{i}.$
\end{fakt}
Forms $\theta ^{i}$ are the forms on $U$ which are dual to the basis $X_{k}$ on $U\subset M_{n}$, i.e. $\theta ^{i}(X_{k})=\delta ^{i}_{\ k}$.

Let $\omega $ be a 1-form of linear connection of the manifold $(M_{n},A_{M}):\ \omega =\omega ^{i}_{\ j}E^{\ j}_{i}$, where $\lbrace E^{\ j}_{i}\rbrace $ is a natural basis of the algebra $\mathfrak{gl}(n,\mathbb{R})$ of a Lie group $GL(n, \mathbb{R})$. Let $\sigma :\ U\rightarrow L(M)$ be a section of the principal bundle $L[M,GL(n,\mathbb{R})]$ over $U\subset M_{n}$, which attaches for every $x\in U$ a linear frame $\lbrace X_{i}(x)\rbrace \ \ (i=1,...,n).$\\
Denote $\omega _{U}:=\sigma_{*}\omega $, where $\omega _{U}$ is a pull-back of the connection form $\omega $ from $L(M)$ bundle onto $U$. $\omega _{U}$ is a 1-form on $U\subset M_{n}$ with values in the algebra $\mathfrak{gl}(n,\mathbb{R}): \ \omega _{U}=\omega _{U\ k}^{\ \ i}E^{\ k}_{i}.$ Let decompose 1-forms $\omega ^{\ \ i}_{U\ k}$ in a natural cobasis $\lbrace dx^{j}=\sigma_{*}\theta ^{j}\rbrace $ of 1-forms on $U:$
$$\omega ^{\ \ i}_{U\ k}:=\Gamma ^{i}_{kj}dx^{j}.$$
$\lbrace \Gamma ^{i}_{kj}\rbrace (i,j,k=1,...,n)$ are the components of the pull-back $\omega ^{\ \ i}_{U\ k}$ of the linear connection $\Gamma $ on $L(M)$ in the local coordinates on the base manifold $M_{n}$. It can be shown, that $\Gamma ^{i}_{kj}$ undergo the following transformation law on the intersection of two local charts on $M_{n}:$
\begin{align}
\Gamma ^{i'}_{k'j'}(P)=\frac{\partial x^{i'}}{\partial x^{i}}(P)\frac{\partial x^{k}}{\partial x^{k'}}(P)\frac{\partial x^{j}}{\partial x^{j'}}(P)\Gamma ^{i}_{kj}(P)+\frac{\partial ^{2}x^{l}}{\partial x^{k'}\partial x^{j'}}(P)\frac{\partial x^{i'}}{\partial x^{l}}(P)
\end{align}
$P\in U\cap V.$ On $U$ we have local coordinates $\lbrace x^{i}\rbrace (i=1,...,n)$ and on $V$ local coordinates $\lbrace x^{i'}\rbrace (i'=1,...,n)$. 

The above formula is the very known transformation law of the linear connection components $\Gamma ^{i}_{kl}$ in the local charts on $(M_{n}, A_{M})$. It is obtained [8] from the transformation law given in Fact 4. \\
Thus from the general global theory of connection (Ehresmann) we have got the local index theory of connection on the differentiable manifold $(M_{n},A_{M})$. This theory was fully developed in the past mainly by J.A.Schouten. It is also possible to show the inverse fact: the local components $\Gamma ^{i}_{kl}$ which satisfy (1) determine a unique linear connection $\Gamma $ in $L(M)$.\\
We had on $L(M)$:
$$\Theta ^{i}=\frac{1}{2}Q^{i}_{kl}\theta ^{k}\wedge \theta ^{l},$$
$$\Omega ^{i}_{j}=\frac{1}{2}R^{i}_{jkl}\theta ^{k}\wedge \theta ^{l}.$$
Let $\sigma :\ U\rightarrow L(M)$ be a local section of $L(M)$ over $U$.\\
We define:
$$\theta ^{i}_{U}:=\sigma _{*}\theta ^{i}(=dx^{i}),$$
$$\Theta ^{\ i}_{U}:=\sigma _{*}\Theta ^{i}=:\frac{1}{2}\widetilde{Q}^{i}_{.kl}dx^{k}\wedge dx^{l},$$
$$\Omega ^{\ \ i}_{U\ j}:=\sigma _{*}\Omega ^{i}_{j}=\frac{1}{2}\widetilde{R}^{i}_{jkl}dx^{k}\wedge dx^{l}.$$
It appears [8], that:
$$\widetilde{Q}^{i}_{.kl}=-\widetilde{Q}^{i}_{.lk}=\Gamma ^{i}_{lk}-\Gamma ^{i}_{kl},$$
$$\widetilde{R}^{i}_{\ jkl}=-\widetilde{R}^{i}_{\ jlk}=\Gamma ^{i}_{\ jl,k}-\Gamma ^{i}_{\ jk,l}+\Gamma ^{i}_{\ mk}\Gamma ^{m}_{\ jl}-\Gamma ^{i}_{\ ml}\Gamma ^{m}_{\ jk}.$$

The above formulas are standard expressions for components of the torsion and curvature tensors in the local charts on the differentiable manifold $(M_{n},A_{M})$.
\begin{uwaga}
\begin{enumerate}
\item For the local forms $\Omega _{U\ j}^{\ \ i}$, $\Theta ^{\ \ i}_{U}$,  $\omega ^{\ \ i}_{U\ k}$, $\theta ^{\ \ i}_{U}$, we have the same relations as on the bundle $L(M)$\\
$\Omega _{U\ j}^{\ \ i}=d\omega ^{\ \ i}_{U\ j}+\omega ^{\ \ i}_{U\ k}\wedge \omega ^{\ \ k}_{U\ j}$\\
$\Theta ^{\ \ i}_{U}=d\theta ^{\ \ i}_{U}+\omega ^{\ \ i}_{U\ k}\wedge \theta ^{\ \ k}_{U}.$\\
\item For the Riemannian connection (and for the pseudoriemannian connection) we have in the local coordinates on the Riemannian (pseudoriemannian) differentiable manifold $(M_{n},\widetilde{g})$: $\sigma _{*}\theta ^{i}=dx^{i}$, $\Theta ^{i}=0\implies \Theta ^{\ \ i}_{U}=0\implies \widetilde{Q}^{i}_{kl}=0$, $\mathcal{D}\widetilde{g}_{lk}=0\equiv dx^{i}\widetilde{g}_{lk;i}=0\equiv dx^{i}\bigl (\widetilde{g}_{lk,i}-\Gamma ^{p}_{\ li}\widetilde{g}_{pk}-\Gamma ^{p}_{\ ki}\widetilde{g}_{lp} \bigr ) $, where $\widetilde{g}_{lk}(x)=\sigma _{*}g_{lk}=g_{lk}\sigma $, $g_{lk}:\ L(M)\rightarrow \mathbb{R}\ \ \frac{n(n+1)}{2}$ of functions on $L(M)$, $\widetilde{g}_{lk}$ is a metric on $U \subset M_{n}$ and $\Gamma ^{i}_{\ kl}=\Gamma ^{i}_{\ lk}=\lbrace ^{i}_{kl}\rbrace =\frac{1}{2}\widetilde{g}^{im}\bigl (\widetilde{g}_{km,l}+\widetilde{g}_{lm,k}-\widetilde{g}_{kl,m}\bigr )$. These formulas are well-known for the local Riemannian geometry on the base $(M_{n},\widetilde{g})$.
\end{enumerate}
\end{uwaga}

\underline{Bianchi identities for Riemannian  (Levi-Civita) connection}\\
\underline{in the local charts on $(M_{n},g)$} and in terms of the Riemannian curvature tensor:\\
These are the identities\\
$\widetilde{R}^{k}_{\ [mrl]}\equiv 0$ (the first) \\
or after extending alternation\\
$\widetilde{R}^{k}_{\ mrl}+\widetilde{R}^{k}_{\ rlm}+\widetilde{R}^{k}_{\ lmr}\equiv 0$.\\
$\widetilde{R}^{k}_{\ l[rs;m]}\equiv 0$ (the second)\\
or \\
$\widetilde{R}^{k}_{\ lrs;m}+\widetilde{R}^{k}_{\ lsm;r}+\widetilde{R}^{k}_{\ lmr;s}\equiv 0$ after extending alternation.

All the above formulas (with missing tylda) are still extensively used in standard books on relativistic theory of gravity.\\

Below we give other calculational formulas very important in this theory.

Let $\rho $ be a representation of the group $GL(n,\mathcal{R})$, i.e. a representation of the group of structure principal fibre budle of linear frames $L(M)$ in the vector space $V=T^{r}_{s}$.

$T^{r}_{s}=\stackrel {r}{\otimes }V^{n}\otimes \stackrel {s}{\otimes }V_{n}^{\ *}$ is a vector space of tensors of the type $(r,s)$ and $\rho :\ GL(n,\mathbb{R})\rightarrow GL(V)$ is a homomorphism of Lie groups $GL(n,\mathbb{R})$ and $GL(V)=\stackrel {r+s}{\otimes }GL(n,\mathbb{R})$.

\begin{definicja}
A pseudotensorial $q$-form of the type $\rho $ on $L(M)$ is a $q$-form $\alpha $ on $L(M)$ such that 
$$(R_{a})_{*}\alpha (X_{1},...,X_{q})=\rho _{a^{-1}}\cdot [\alpha (X_{1},...,X_{q})],$$
$a\in GL(n, \mathbb{R})$, $\ X_{1},...,X_{q}\in T_{a}[L(M)]$.
\end{definicja}

\begin{definicja}
$hor \cdot \alpha :=\alpha \cdot h(X_{1},...,X_{q})=\alpha (hX_{1},...,hX_{q}).$
\end{definicja}

\begin{definicja}
A form $\alpha $ is tensorial ($\equiv $ horizontal) $\iff $ $hor\cdot \alpha \equiv \alpha $.
\end{definicja}

Tensorial forms on $L(M)$ with values in $T^{r}_{s}$ are used in teoretical physics to describe material fields. In the case of a manifold $(M_{n},A_{m})$ the space $V_{n}$ is $T_{x}(M)$ and $V^{*}_{n}=T_{x}^{*}(M)$. 

\begin{definicja}
An exterior covariant differential of the tensorial $q$-form of the type $\rho $ on $L(M)$ is a tensorial $(q+1)$-form of the type $\rho $ denoted by $\mathcal{D}\alpha $, where 
$$\mathcal{D}\alpha :=hor\cdot d\alpha =d\alpha (hX_{1}, hX_{2},..., hX_{q}, hX_{q+1}).$$
\end{definicja}

Hereafter we limit ourselves to the tensorial $q$-forms on $L(M)$ with values in $T^{r}_{s}=\stackrel{r}{\otimes }T_{x}(M_{n})\otimes \stackrel{s}{\otimes }T_{x}^{\ *}(M_{n})$.

Let $\sigma :\ U\rightarrow L(M)$ be a local section of the bundle such that $L(M):\ \pi \cdot [\sigma (x)]=x$, where
$x\in U\subset M_{n}$, $\sigma (x)=\lbrace  x_{1}(x),...,x_{n}(x), X_{1}(x),...,X_{n}(x)\rbrace \in L(M)$, $\lbrace X_{1}(x),...,X_{n}(x)\rbrace $ is a linear basis in $T_{x}(M)$.

Then $\Lambda =\sigma _{*}\alpha :=\alpha \cdot \sigma $ is a tensorial $q$-form on $U\in M_{n}$ with values in $V=T^{r}_{s}=\stackrel {r}{\otimes }T_{x}(M_{n})\otimes \stackrel {s}{\otimes }T_{x}^{\ *}(M_{n}),$ where $x\in U\subset M_{n}.$

This form can be written in the local chart on $U\subset M_{n}$ in the form
$$\stackrel {q}{\Lambda }=\Lambda _{i_{1}...i_{s}\ \ \ \ \ k_{1}...k_{q}\atop k_{1}<k_{2}<...<k_{q}}^{\ \ \ \ \ j_{1}...j_{r}}(x)\partial_{j_{1}}\otimes ...\otimes \partial _{j_{r}}\otimes dx^{i_{1}}\otimes ...\otimes dx^{i_{s}}\otimes dx^{k_{1}}\wedge ... \wedge dx^{k_{q}}.$$

Mappings $\lbrace \partial_{j_{1}}\otimes ...\otimes \partial _{j_{r}}\otimes dx^{i_{1}}\otimes ...\otimes dx^{i_{s}} \rbrace$ form the natural basis in $T^{r}_{s}=\stackrel {r}{\otimes }T_{x}(M_{n})\otimes \stackrel {s}{\otimes }T_{x}^{\ *}(M_{n})$ and mappings $(dx^{k_{1}}\wedge ... \wedge dx^{k_{q}})$ form the natural basis $q$-form on $U\subset M_{n}$.

In practice we write the above tensorial q-form of the type $(r,s)$ as follows
$$\stackrel {q}{\Lambda }=\stackrel {q}{\Lambda }_{i_{1}...i_{s}}^{\ \ \ \ \ j_{1}...j_{r}}B_{j_{1}...j_{r}}^{\ \ \ \ \ i_{1}...i_{s}},$$
where $B_{j_{1}...j_{r}}^{\ \ \ \ \ i_{1}...i_{s}}= \partial_{j_{1}}\otimes ...\otimes \partial _{j_{r}}\otimes dx^{i_{1}}\otimes ...\otimes dx^{i_{s}}, $ remembering that every component of $\stackrel {q}{\Lambda }_{i_{1}...i_{s}}^{\ \ \ \ \ j_{1}...j_{r}}$ is $q$-form 
$$\stackrel {q}{\Lambda }_{i_{1}...i_{s}}^{\ \ \ \ \ j_{1}...j_{r}}=\sum_{k_{1}<k_{2}<...<k_{q}}\Lambda_{i{1}...i_{s}\ \ \ \ \ k_{1}...k_{q}}^{\ \ \ \ \ j_{1}...j_{r}}dx^{k_{1}}\wedge ...\wedge dx^{k_{q}}$$. 

We define the exterior differential $d\stackrel {q}{\Lambda }$
$$d\stackrel {q}{\Lambda }=d\stackrel {q}{\Lambda }_{i_{1}...i_{s}}^{\ \ \ \ \ j_{1}...j_{r}}\otimes B_{j_{1}...j_{r}}^{\ \ \ \ \ i_{1}...i_{s}},$$
where $d\stackrel {q}{\Lambda }_{i_{1}...i_{s}}^{\ \ \ \ \ j_{1}...j_{r}}$ means Cartan derivative of the $q$-form $$\stackrel {q}{\Lambda }_{i_{1}...i_{s}}^{\ \ \ \ \ j_{1}...j_{r}}=\sum_{k_{1}<k_{2}<...<k_{q}}\Lambda_{i{1}...i_{s}\ \ \ \ \ k_{1}...k_{q}}^{\ \ \ \ \ j_{1}...j_{r}}dx^{k_{1}}\wedge ...\wedge dx^{k_{q}}$$. 

The action $d$ is linear and $d\cdot d\stackrel {q}{\Lambda }\equiv 0$ for $\stackrel {q}{\Lambda }.$ But it appears that $d\stackrel {q}{\Lambda }$ is not the tensorial $(q+1)$-form with values in $T^{r}_{s}=\stackrel {r}{\otimes }T_{x}(M_{n})\otimes [\stackrel {s}{\otimes }T^{\ *}_{x}(M_{n})]$, $x\in M_{n}$.

\underline{An exterior covariant differential $\mathcal{D}\stackrel {q}{\Lambda }$} is the tensorial $(q+1)$-form on $U\subset M_{n}$ with values in $T^{r}_{s}$.

One can show that [9]
$$\mathcal{D}\stackrel {q}{\Lambda }=\mathcal{D}\stackrel {q}{\Lambda }_{i_{1}...i_{s}}^{\ \ \ \ \ j_{1}...j_{r}}\otimes B_{j_{1}...j_{r}}^{\ \ \ \ \ i_{1}...i_{s}},$$
where $\mathcal{D}\stackrel {q}{\Lambda }_{i_{1}...i_{s}}^{\ \ \ \ \ j_{1}...j_{r}}=d\stackrel {q}{\Lambda }_{i_{1}...i_{s}}^{\ \ \ \ \ j_{1}...j_{r}}+\omega ^{j_{1}}_{\ \ p}\wedge \stackrel {q}{\Lambda }_{i_{1}...i_{s}}^{\ \ \ \ \ pj_{2}...j_{r}}+\omega ^{j_{2}}_{\ \ p}\wedge \stackrel {q}{\Lambda }_{i_{1}...i_{s}}^{\ \ \ \ \ j_{1}pj_{3}...j_{r}}+...+\omega ^{j_{r}}_{\ \ p}\wedge \stackrel {q}{\Lambda }_{i_{1}...i_{s}}^{\ \ \ \ \ j_{1}j_{2}...p}-\omega ^{p}_{\ \ i_{1}}\wedge \stackrel {q}{\Lambda }_{pi_{2}...i_{s}}^{\ \ \ \ \ j_{1}...j_{r}}-\omega ^{p}_{\ \ i_{2}}\wedge \stackrel {q}{\Lambda }_{i_{1}p...i_{s}}^{\ \ \ \ \ j_{1}...j_{r}}-...-\omega ^{p}_{\ \ i_{s}}\wedge \stackrel {q}{\Lambda }_{i_{1}i_{2}...p}^{\ \ \ \ \ j_{1}...j_{r}}$.

By $\omega ^{i}_{\ \ p}$ one should understand the local forms $\omega ^{\ i}_{U\ \ \ p}=\sigma _{*}\omega ^{i}_{\ \ p}$.\\ 
In the natural cobasis on $U\subset M_{n}$ we have $\omega ^{i}_{\ \ k}=\Gamma ^{i}_{\ \ kl}dx^{l}$.

In nonholonomic Lorentzian coreper $(\vartheta ^{I}):\ g=\eta _{IK}\vartheta ^{I}\otimes \vartheta ^{K}$ we put $\omega ^{I}_{\ \ K}=\gamma ^{I}_{\ \ KL}\vartheta ^{L}$.

The important relation between $\gamma ^{I}_{KL}$ and $\Gamma ^{i}_{kl}$ for a metric connection $\omega ^{i}_{\ \ p}$ is the following:
$$\Gamma ^{i}_{\ \ kl}(x)=h^{i}_{\ A}(x)\gamma ^{A}_{\ BC}(x)h^{B}_{\ k}h^{C}_{\ l}(x)+h^{i}_{\ A}(x)\partial _{k}h^{A}_{\ l}(x),$$
where matrices $(h^{I}_{\ a})$ and $h^{a}_{\ B}$ are defines as follows
$$\vartheta ^{I}=h^{I}_{\ a}(x)dx^{a}$$
$$h^{a}_{\ B}(x)h^{B}_{\ b}(x)=\delta ^{a}_{b}\equiv h^{a}_{\ B}(x)h^{D}_{\ a}(x)=\delta ^{D}_{B}.$$

The inverse relation has the form 
$$\gamma ^{A}_{\ \ BC}=h^{g}_{\ C}h^{b}_{\ B}h^{A}_{\ a}\Gamma ^{a}_{\ bg}-h^{\gamma }_{\ C}h^{\beta }_{\ B}\partial _{\beta }(h^{A}_{\ \gamma }).$$

For components of a tensor-valued form one has simpler relations, e.g. for curvature components we have 
$$R_{ABCD}=h^{a}_{\ A}h^{b}_{\ B}h^{c}_{\ C}h^{d}_{\ D}R_{abcd},$$
$$R_{abcd}=h^{A}_{\ a}h^{B}_{\ b}h^{C}_{\ c}h^{D}_{\ d}R_{ABCD}.$$

If $\alpha $ is a tensorial 0-form with values in $T^{r}_{s}$, then $\Lambda =\sigma _{*}\alpha $ is a tensorial field of the type  $(r,s)$ on $U\subset M_{n}$, i.e. $\Lambda $ is a section of tensor bundle of the type $(r,s)$. 

Then in the natural basis and cobasis we have on $U\subset M_{4}$
$$\Lambda =\Lambda _{i_{1}...i_{s}}^{\ \ \ \ \ j_{1}...j_{r}}\otimes B_{j_{1}...j_{r}}^{i_{1}...i_{s}}.$$
In this case
$$\mathcal{D}\Lambda =\mathcal{D}\Lambda _{i_{1}...i_{s}}^{\ \ \ \ \ j_{1}...j_{r}}\otimes B_{j_{1}...j_{r}}^{i_{1}...i_{s}},$$
where $\mathcal{D}\Lambda _{i_{1}...i_{s}}^{\ \ \ \ \ j_{1}...j_{r}}(x)=d\Lambda _{i_{1}...i_{s}}^{\ \ \ \ \ j_{1}...j_{r}}(x)+\omega ^{j_{1}}_{\ p}\Lambda _{i_{1}...i_{s}}^{\ \ \ \ \ pj_{2}...j_{r}}+...+\omega ^{j_{r}}_{\ p}\Lambda _{i_{1}...i_{s}}^{\ \ \ \ \ j_{1}j_{2}...p}-\omega ^{l}_{\ t_{1}}\Lambda _{li_{2}...i_{s}}^{\ \ \ \ \ j_{1}...j_{r}}-...-\omega ^{l}_{\ i_{s}}\Lambda _{i_{1}i_{2}...l}^{\ \ \ \ \ j_{1}...j_{r}}$.

If we decompose the connection 1-form $(\omega ^{i}_{k})$ in the natural cobasis $U\subset M_{4}$
$$\omega ^{i}_{k}=\Gamma ^{i}_{kl}dx^{l}$$
then we obtain that
$$\mathcal{D}\Lambda _{i_{1}...i_{s}}^{\ \ \ \ \ j_{1}...j_{r}}=\Lambda _{i_{1}...i_{s}\ \ \ \ \ \ ;k}^{\ \ \ \ \ j_{1}...j_{r}}dx^{k}.$$
We can see that $\mathcal{D}\Lambda _{i_{1}...i_{s}}$ are components of the absolute differential of the tensorial field $\Lambda $, and $\Lambda _{i_{1}...i_{s}}^{\ \ \ \ \ j_{1}...j_{r}};_{k}$ are components of the covariant derivative of this field.
One can read from the last formula that 
\begin{align}
\Lambda _{i_{1}...i_{s}}^{\ \ \ \ \ j_{1}...j_{r}};_{l}=&\Lambda _{i_{1}...i_{s}}^{\ \ \ \ \ j_{1}...j_{r}},_{l}+\Gamma ^{j_{1}}_{pl}\Lambda _{i_{1}...i_{s}}^{\ \ \ \ \ pj_{2}...j_{r}}+...+\Gamma ^{j_{r}}_{pl}\Lambda _{i_{1}...i_{s}}^{\ \ \ \ \ j_{1}...p} \notag
\\&-\Gamma ^{p}_{i_{1}l}\Lambda _{ki_{2}...i_{s}}^{\ \ \ \ \ j_{1}...j_{r}}-...-\Gamma ^{p}_{i_{s}l}\Lambda _{i_{1}...p}^{\ \ \ \ \ j_{1}...j_{r}}, \notag
\end{align}

$\Lambda _{i_{1}...i_{s}}^{\ \ \ \ \ j_{1}...j_{r}};_{k}\in T^{r}_{s+1}.$

If $\alpha $ is an ordinary Cartan q-form with values in $\mathfrak{R}$, then $\Lambda =\sigma _{*}\alpha $ is an ordinary Cartan q-form on $U\subset M_{4}$:
$$\Lambda =\sum _{i_{1}<i_{2}<...<i_{q}}\Lambda _{i_{1}...i_{q}}dx^{i_{1}}\wedge ...\wedge dx^{i_{q}}.$$
Then $\mathcal{D}\Lambda =d\Lambda $, where $d$ denotes Cartan exterior differential. We have
\begin{align}
d \stackrel {q}{\Lambda }=\sum_{i_{1}<i_{2}<...<i_{q}}d\Lambda _{i_{1}...i_{q}}\wedge dx^{i_{1}}\wedge ...\wedge dx^{i_{q}}. 
\end{align}
$d\Lambda _{i_{1}...i_{q}}$ denotes here ordinary differential of the components $\Lambda _{i_{1}...i_{q}}$ of the q-form $\stackrel {q}{\Lambda }$. From (2) one can obtain the formula for practical calculations [9]
\begin{align}
d \stackrel {q}{\Lambda }=\sum_{i_{0}<i_{1}<...<i_{q}}\Bigl (\sum^{q}_{\alpha =0}(-1)^{\alpha }\partial _{i_{\alpha }}\Lambda _{i_{0}i_{1}...\hat {i}_{\alpha}...i_{q}(x)}\Bigr )dx^{i_{0}}\wedge ...\wedge dx^{i_{q}} . 
\end{align}
"$\hat {}$" over the index means that this index should be left.

From the above formulas we can see that the exterior covariant differential $\mathcal{D}$ is the most general differentiation on the manifold. Namely, the absolute differential of the tensorial field $\mathcal{D}$ and the external differential $d$ of the external Cartan form are the particular examples of this differentiation.\\
All formulas given above are used in standard computations in General Relativity (generally: in theoretical physics).

\end{document}